\documentclass[aps,prc,twocolumn,showpacs,amsmath,amssymb,preprintnumbers]{revtex4}

\usepackage[10pt]{type1ec}  
\usepackage[T1]{fontenc}
\usepackage{CJKutf8}
\usepackage[overlap, CJK]{ruby}
\usepackage{CJKulem}
\newenvironment{SChinese}{%
  \CJKfamily{gbsn}%
  \CJKtilde
  \CJKnospace}{}

\usepackage[bookmarks=false]{hyperref}

\usepackage{latexsym} 
\usepackage{graphicx}       
\usepackage{color}
\usepackage{bm}   

\newcommand{\al}{\alpha}

\newcommand{\de}{\delta}

\newcommand{\ep}{\varepsilon}
\newcommand{\eps}{\epsilon}

\renewcommand{\th}{\theta}   

\newcommand{\om}{\omega}
\newcommand{\Om}{\Omega}

\newcommand{\beq}{\begin{equation}}
\newcommand{\eeq}{\end{equation}}
\newcommand{\ba}{\begin{array}}
\newcommand{\ea}{\end{array}}
\newcommand{\bea}{\begin{eqnarray}}
\newcommand{\eea}{\end{eqnarray}}
\newcommand{\bi}{\begin{itemize}}  
\newcommand{\ei}{\end{itemize}}
\newcommand{\ben}{\begin{enumerate}} 
\newcommand{\een}{\end{enumerate}}
\newcommand{\bc}{\begin{center}}
\newcommand{\ec}{\end{center}}

\newcommand{\txt}{\textstyle}
\newcommand{\dsp}{\displaystyle}
\newcommand\eqn[1]{(\ref{#1})}      

\newcommand{\half} {{\txt \frac{1}{2}}}

\begin{document}

\title{Phase conversion dissipation in multicomponent compact stars}
\author{Mark G. Alford, Sophia Han
(\begin{CJK}{UTF8}{}\begin{SChinese}韩 君\end{SChinese}\end{CJK}), Kai Schwenzer
}

\affiliation{Physics Department, Washington University,
St.~Louis, Missouri~63130, USA}

\begin{abstract}
We propose a mechanism for the damping of density oscillations in
multicomponent compact stars. The mechanism is the periodic conversion
between different phases, i.e.,~the movement of the interface between
  them, induced by pressure oscillations in the star. The damping grows
nonlinearly with the amplitude of the oscillation.
We study in detail the case of
r-modes in a hybrid star with a sharp interface, and we find that this
mechanism is powerful enough to saturate the r-mode at 
very low saturation amplitude, of order $10^{-10}$, and
is therefore likely to be the dominant r-mode saturation mechanism in 
hybrid stars with a sharp interface.
\end{abstract}

\date{11 May 2015} 



\maketitle

\section{Introduction}

The damping of mechanical oscillations of compact stars is a promising 
signature of the phases of dense matter in their interior. The damping of density perturbations, described locally by the bulk viscosity, 
is particularly important since it has been shown to vary greatly
between different phases \cite{Sawyer:1989dp,Haensel:1992zz,Madsen:1992sx,Haensel:2000vz,Haensel:2001mw,Haensel:2001em,Jones:2001ya,Alford:2006gy,Manuel:2007pz,Alford:2007rw,Mannarelli:2009ia,Wang:2010ydb,Schwenzer:2012ga,Manuel:2013bwa}. In addition to the damping properties of bulk phases, the boundary between different phases can also be relevant for dissipation. A well-known example is Ekman layer damping due to shear forces at the boundary between a fluid and a solid phase \cite{Lindblom:2000gu}. Here we propose a dissipation mechanism that stems from the fact that pressure oscillations can cause the interface between two  phases to move back and forth, as the two phases are periodically converted into each other. If the finite rate of this conversion produces a phase lag between the pressure oscillation and the position of the interface, energy will be dissipated in each cycle. We study the resultant damping for the case of a hybrid star with a sharp interface between the quark core and the hadronic mantle, where the dissipation is due to quark-hadron burning at the interface. However, the mechanism is generic and could be relevant for any star with an internal interface between phases of different energy density.

Unstable global oscillation modes \cite{Friedman:1978hf} are of particular interest since they arise spontaneously and grow until stopped by some saturation (nonlinear damping) mechanism. For neutron stars, the most important example is r-modes, \cite{Andersson:1997xt,Andersson:2000mf} since they are unstable in typical millisecond pulsars unless sufficient damping is present.
Several mechanisms for the saturation of the growth of unstable r-modes have been proposed \cite{Lindblom:2000az,Arras:2002dw,Bondarescu:2007jw,Alford:2011pi,Bondarescu:2013xwa,Haskell:2013hja}. Although bulk viscosity has
a nonlinear ``suprathermal'' regime \cite{Madsen:1992sx,Alford:2010gw,Alford:2011df}, it has been found that this becomes relevant only at very high amplitudes, and is probably pre-empted by some other stronger mechanism \cite{Alford:2011pi}.
In this paper we show that dissipation due to hadron-quark burning could well be the dominant r-mode saturation mechanism in hybrid stars. The dissipation is vanishingly small at infinitesimal amplitude, but becomes very strong as the amplitude increases. (For similar behavior in a different context, see Ref.~\cite{Haskell:2013hja}). This strong dissipation saturates unstable r-modes in compact stars with a sufficiently large core at amplitudes that are orders of magnitude below those provided by any other known saturation mechanism. We give a simple analytic prediction for the saturation amplitude, and find that it can be as low as $\alpha_{{\rm sat}}\lesssim10^{-10}$ for conditions present in observed pulsars. 

\section{schematic model for the dissipation due to phase transformation}
\label{sec:piston}

\subsection{Two phases in a cylinder}

As a step towards an analysis of the dissipation due to phase conversion in an inhomogeneous multicomponent star, we now construct a simplified version of the interface between different layers in a gravitationally bound system.
We calculate the energy dissipated
in this system when it is subjected to periodic compression and rarefaction.

\begin{figure}[htb]
\includegraphics[width=\hsize]{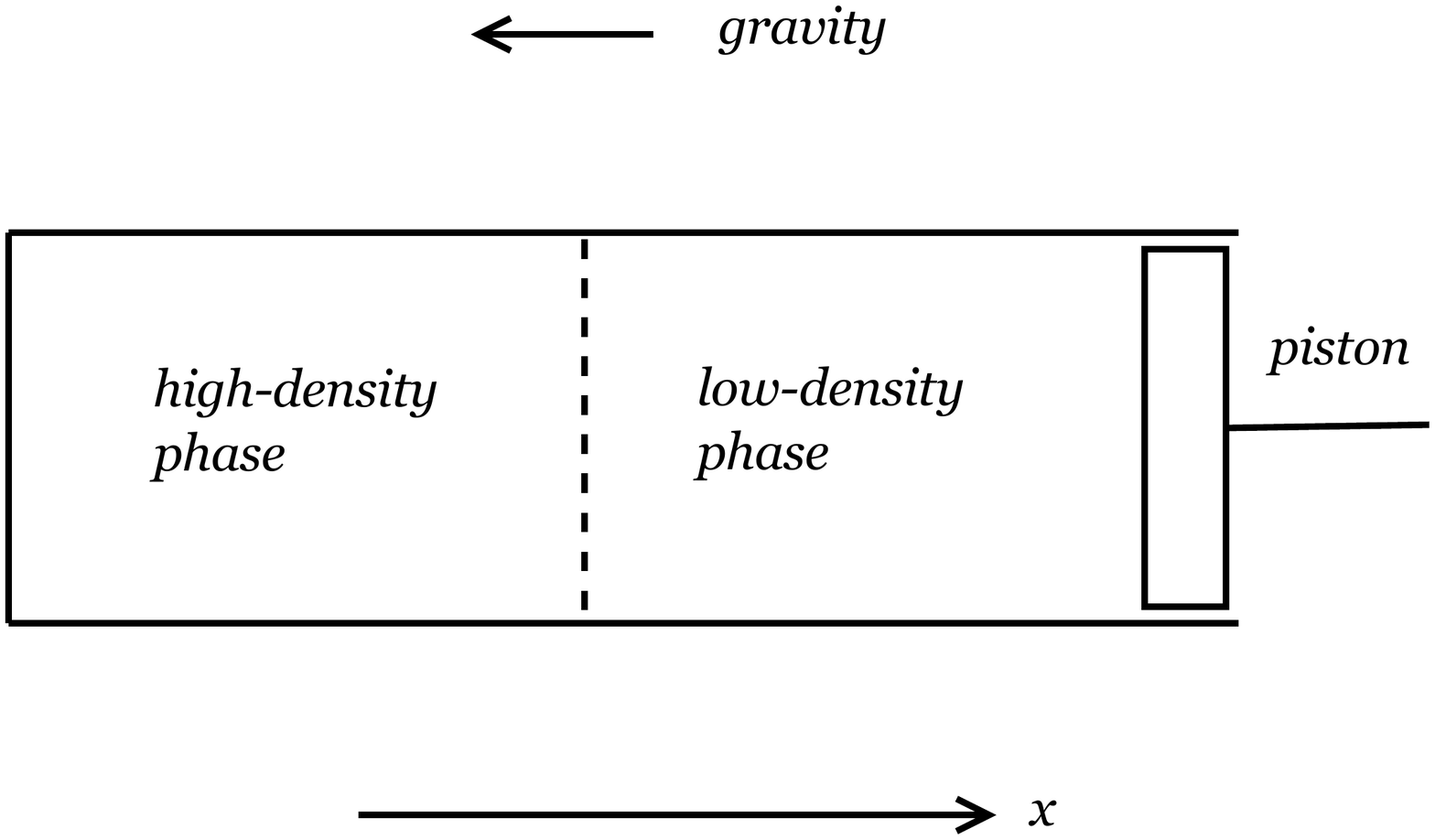}
\caption{
Schematic model: two incompressible phases in a cylinder with
piston, in a gravitational field. An oscillation of the external pressure
on the piston leads to interconversion of the
two phases, and hence movement of the piston.
\label{fig:piston_diag}
}
\end{figure}

Our schematic model system involves two incompressible phases, characterized by different densities of a conserved particle species.
We assume there is a first-order pressure-induced phase transition, so 
the phases are separated by a sharp interface (``the phase boundary'') 
which, in long-term equilibrium, is at the critical pressure $p_{\rm crit}$, and that there are
processes that can convert each phase into the other at some finite rate.
We consider a cylinder containing both phases
in a homogeneous (Newtonian) gravitational field, with a piston which can be moved parallel to the direction of the field (Fig.~\ref{fig:piston_diag}). The high-density phase is deeper in the gravitational potential than the low-density phase. The field produces a pressure gradient in the cylinder, which can be shifted by moving the piston. This will cause the equilibrium position of the interface to shift, but, crucially, depending on the
speed of the conversion process, it may take some
time for the interface to move to its new equilibrium position.
This causes the response of the system (its volume or density) to 
lag behind the externally applied force, resulting in dissipation.
To calculate the energy dissipated per cycle, we simply calculate the net
$p\,\mathrm{d}V$ work done by the piston in one cycle.

We assume that the
 equation of state (EoS) of the two phases is linear,
\beq
p(\mu) = \left\{\!
\begin{array}{l@{\quad}l}
\mu \, n_{\rm L}-\ep_{\rm L} & \mbox{(low-density phase)} \\
\mu \, n_{\rm H}-\ep_{\rm H} & \mbox{(high-density phase)}
\end{array}
\right. \ ,
\label{eqn:EoShyb}
\eeq
where $\mu$ is the chemical potential for the conserved particle number and the two incompressible phases have fixed
particle number densities $n_{\rm L}$ and $n_{\rm H}$ and 
fixed energy densities $\eps_{\rm L}$ and $\eps_{\rm H}$.
Later we use the fact that this is a valid approximation
for any EoS, as long as the pressure oscillations
are small enough.

\begin{figure}[htb]
\includegraphics[width=\hsize]{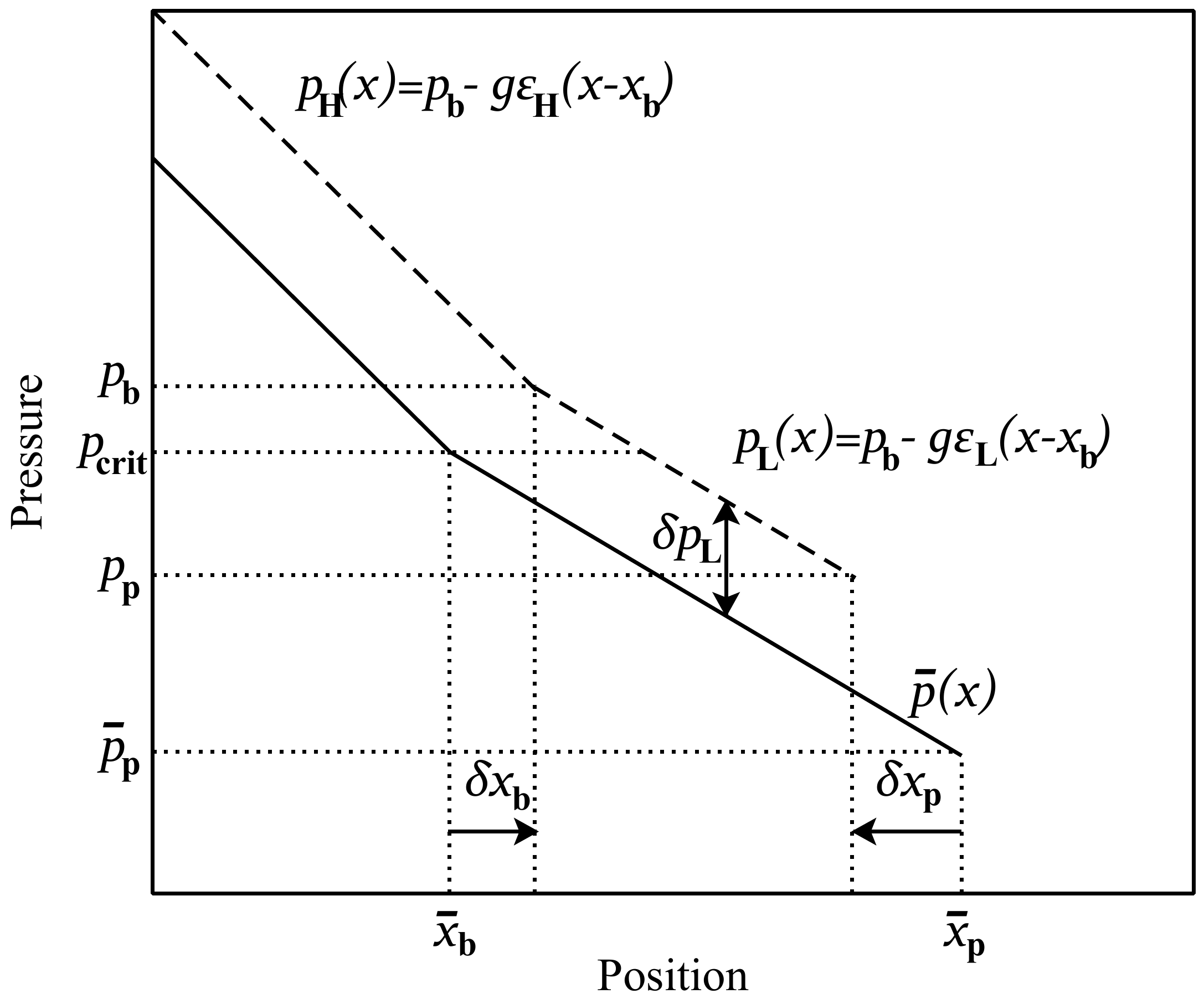}
\caption{
Pressure gradients in the cylinder of Fig.~\ref{fig:piston_diag}.
The solid line $\bar p(x)$ is the pressure profile in long-term equilibrium.
The dashed line is a snapshot of the system at a moment when
the piston has moved inward a distance $\de x_{\rm p}$, the
pressure everywhere has risen, and the phase boundary 
has moved out a distance $\de x_{\rm b}$ as the low density phase
in part of the region with $p>p_{\rm crit}$ 
has converted to the high-density phase.
}
\label{fig:p_r}
\end{figure}

In a Newtonian gravitational field the pressure is 
a function of $x$ determined by
 \beq
\frac{d p}{d x} = -g\,\ep \ ,
\label{eqn:p_grav}
\eeq
where $g$ is the gravitational acceleration, 
assumed to be independent of $x$.
Equation~\eqn{eqn:p_grav} has a simple solution 
where the pressure varies linearly with $x$, with
a fixed gradient $g\,\ep$ in each phase
(see  Fig.~\ref{fig:p_r})
\beq
p(x) = \left\{\!
\begin{array}{l@{\qquad}l}
p_{\rm b}-g\ep_{\rm H}(x-x_{\rm b}) & x<x_{\rm b} \\
p_{\rm b}-g\ep_{\rm L}(x-x_{\rm b}) & x\geqslant{x_{\rm b}}
\end{array}
\right. \ ,
\label{eqn:p_r}
\eeq
where $x_{\rm b}$ is the position of the interface between the two phases (``the boundary'') and $p_{\rm b}$ is the pressure at the boundary.
In long-term equilibrium, the boundary settles at its ``ideal'' position,
where $p_{\rm b}$ is $p_{\rm crit}$ (see below).

\subsection{External pressure oscillation}

Assume that the external pressure on the piston varies
periodically.
When the pressure is high, part of the low-density phase is driven to a pressure above $p_{\rm crit}$ and starts to convert into the high-density phase, and vice versa during rarefaction. 
The pressure at any given location
and the position of the phase boundary therefore vary in time,
\bea
p(x,t) &=& \bar p(x)+\delta p(x,t) \ , \\
x_{\rm b}(t) &=& \bar{x}_{\rm b}+\delta x_{\rm b}(t) \ ,
\label{eqn:x_Q}
\eea
where $\bar{x}_{\rm b}$ is the equilibrium position of the boundary
and $\bar p(x)$ is the pressure profile in long-term equilibrium.
The position of the boundary at a given moment 
depends on the previous compression history
and the phase conversion rate, 
and we expect that because of the finite rate of conversion between the
two phases the oscillation of the boundary can be out of
phase with the oscillation of the pressure, and this will lead to
dissipation via net $p\,\mathrm{d}V$ work being done in each cycle.

To calculate the dissipation, we need to relate the movement of the boundary to
the applied pressure oscillation. We assume that the pressure in the low-density phase oscillates harmonically with amplitude $\Delta p_{\rm L}$ and frequency $\omega$, so $\delta p_{\rm L}(t)=\Delta p_{\rm L}\sin(\omega t)$.

In equilibrium, the piston is at $\bar{x}_{\rm p}$ with pressure $\bar{p}_{\rm p}$. As part of the pressure oscillation the piston moves
\beq
x_{\rm p}(t) = \bar{x}_{\rm p}+\delta x_{\rm p}(t) \, ,
\label{eqn:r1_t}
\eeq
and the pressure at the piston is
\beq
p_{\rm p}(t) = \bar{p}_{\rm p}-g\ep_{\rm L}\delta x_{\rm p}(t)+\Delta p_{\rm L}\sin(\omega t) 
\label{eqn:p1_t}
\eeq
The movement of the piston and the movement of the phase boundary are connected by particle number conservation inside the cylinder. The total particle number is $N_{\rm tot}= 
(x_{\rm b}(t)n_{\rm H} + (x_{\rm p}(t)-x_{\rm b}(t))n_{\rm L})\,S$ where $S$ is the cross-sectional area. 
Particle number conservation $\delta N_{\rm tot}=0$ gives
\beq
x_{\rm p}(t) = \bar{x}_{\rm p}+\left(1-\frac{n_{\rm H}}{n_{\rm L}}\right)\delta{x}_{\rm b}(t) \ .
\label{eqn:r1_t_xq}
\eeq
We can now express the $p\,\mathrm{d}V$ work done by the piston in one cycle
in terms of the movement of the boundary $\de x_{\rm b}$
induced by the pressure oscillation.
In later sections we will study how the boundary moves, expressing
it as a function of the speed of the phase conversion process.
First, however, we define a useful concept, the ``ideal boundary.''

\subsection{Ideal position of the phase boundary}

In discussing the motion of the boundary it is convenient to
define an ``ideal position'' of the boundary, $x_{\rm ib}$.
Since we are assuming that an external force imposes a specified
time dependence of the pressure in the low-density phase, it is natural
to define the ideal boundary at time $t$ to be the position the boundary
would reach if we held $\delta p_{\rm L}$ fixed at its current value and waited for
phase conversion processes
to equilibrate. Thus $x_{\rm ib}(t)$ is the solution
of $p_{\rm L}(x_{\rm ib},t)=p_{\rm crit}$.
Unlike the actual boundary, the ideal boundary is determined simply by the instantaneous value of the applied pressure, with no dependence on previous history or conversion rate. The position of the ideal boundary therefore oscillates in phase with the pressure,
\beq 
\delta x_{\rm ib}(t) = \Delta x_{\rm ib} \sin(\omega t) \; , \; \Delta x_{\rm ib} = \frac{\Delta p_{\rm L}}{g\ep_{\rm L}} ,
\label{eqn:ideal_wall}
\eeq
and its velocity is $90^\circ$ out of phase with its position
\beq
v_{\rm ib} (t)= v_{\rm ib}^{\rm max} \cos (\omega t) \; , \; v_{\rm ib}^{\rm max}= \omega \Delta x_{\rm ib}=  \frac{\omega \Delta p_{\rm L}}{g\ep_{\rm L}} \ .
\label{eqn:vib}
\eeq

For a harmonic pressure oscillation in the low-density phase the ideal boundary moves harmonically. Note, however, that because of the discontinuity in energy density on the phase boundary, the pressure oscillation cannot be simultaneously harmonic in both phases. It is also worth noting that according to our definition
$x_{\rm ib}(t)$ is not in general the place in the cylinder where the pressure at time $t$ is $p_{\rm crit}$: These locations only coincide if the real phase boundary occurs where the pressure is above $p_{\rm crit}$.

\subsection{Energy dissipation in one cycle}
\label{sec:energy_1cycle}

The net $p\,\mathrm{d}V$ work
done by the piston in one cycle ($0\leqslant t<\tau$, $\tau=2\pi/\omega$) is
\beq
W = -S\int_0^\tau p_{\rm p}(t) \frac{\mathrm{d}x_{\rm p}(t)}{\mathrm{d}t}\,\mathrm{d}t,
\label{eqn:W_0}
\eeq
where $p_{\rm p}$ is the pressure at the location of the piston, which
is determined by the applied oscillation of the piston, and
$x_{\rm p}$ is the position of the piston. The piston's 
position depends on the
movement of the boundary $\de x_{\rm b}(t)$ \eqn{eqn:r1_t_xq},
so from Eq.~\eqn{eqn:W_0} the energy dissipation in one cycle is 
\beq
\ba{rcl}
W &=& \dsp S\left(\frac{n_{\rm H}}{n_{\rm L}}-1\right) \left( \int_0^\tau \bar{p}_{\rm p}\frac{\mathrm{d}\delta x_{\rm b}(t)}{\mathrm{d}t}\,\mathrm{d}t \right. \\[3ex]
&&\left.\qquad \qquad\quad \dsp + \int_0^\tau \Delta{p_{\rm L}}\sin(\omega t)\frac{\mathrm{d}\delta x_{\rm b}(t)}{\mathrm{d}t}\,\mathrm{d}t \right).
\ea
\label{eq:W1_W2} 
\eeq
The movement of the boundary is constrained by the detailed physics of the conversion process, which
determines how fast it can move at any given moment.
For quark-hadron conversion we will see that it obeys a differential equation
which expresses the fact that the boundary's maximum velocity depends
on how far out of equilibrium the boundary is, and whether
it needs to move inward or outward to reach equilibrium.
In effect, the real boundary is always chasing the ideal boundary (which is its
long-run equilibrium position), while the ideal boundary
is a moving target, its sinusoidal movement linearly related to the
applied pressure oscillation; see Eqs.~(\ref{eqn:p1_t}) and (\ref{eqn:ideal_wall}).

In Eq.~\eqn{eq:W1_W2} we see that the dissipation vanishes if the two phases have equal densities,
since then the movement of the phase boundary does not change the volume
of the system, so there is no associated $p\,\mathrm{d}V$ work. 

To derive the dissipation we assumed that the pressure oscillation in the low-density phase is harmonic. Had we instead assumed that the pressure oscillation in the high-density phase is harmonic, then the energy dissipation would be slightly bigger, with a difference of $\Delta W/W\simeq\Delta \ep/\ep_{\rm L}$, where $\Delta \ep$ is the energy density discontinuity at the interface.

\section{R-mode damping}
\label{sec:rmode}

We can now calculate the damping of a global oscillation mode in a
hybrid star resulting from the phase conversion mechanism. A comprehensive analysis of this problem requires the detailed density oscillation of the global mode in a star with multiple components separated by density discontinuities due to first-order phase transitions. So far the profiles for global oscillation modes have not been obtained for such a realistic model of a compact star. We therefore estimate the dissipation from a piecewise model for the mode profile, using the known form for a homogeneous star on either side of the phase boundary. We estimate the error due to this simplified procedure below.
Although the amplitude of the mode's density and pressure oscillation varies 
from place to place in the star, the simple model sketched 
in Sec.~\ref{sec:piston}
then applies locally for sufficiently small volume elements containing the interface between the two phases and the entire range over which it moves in response to the oscillation. 

Here we study the case of r-modes because
they are unstable and a sufficiently powerful damping mechanism is required
to ensure that they saturate at a low enough amplitude so that they
do not spin down the quickly rotating compact stars that
we observe.
The Newtonian result for the energy density fluctuation of an m=2 r-mode to leading order in the rotational frequency of the star $\Omega$ is \cite{Alford:2010fd}
\beq
\frac{\delta \ep}{\bar \ep}=\sqrt{\frac{8}{189}} \alpha AR^{2}\Omega ^{2} \left (\frac{r}{R}\right )^{3} {\rm Re} \left[ Y_{3}^{2}(\theta, \phi)e^{i\omega t}\right],
\label{eqn:de_e_rmode}
\eeq
where $\delta \ep=\ep-\bar{\ep}$, $Y_{3}^{2}(\theta, \phi)=\frac{1}{4}\sqrt{\frac{105}{2\pi}} e^{2i\phi}\sin^2\!\theta\cos\theta$, and $A$ is the inverse speed of sound squared
\beq
A\equiv\frac{\partial \ep}{\partial p}
\label{eqn:c_inv}
\eeq
evaluated at equilibrium. $R$ is the radius of the star, $\alpha$ is the dimensionless mode amplitude, and $\om$ is the r-mode frequency $\om = \frac {2}{3}\Om$.

The r-mode involves flows that are dominantly angular rather than radial.
At any moment there is higher pressure in some regions of solid angle
in the star, and lower pressure in other regions. 
This means that globally the fraction of high or low pressure phase does
not change much over time. However, an r-mode will still lead to
conversion between the phases, since the low and high pressure 
regions are kilometers apart, so the gradients of pressure and density
in the angular directions are extremely small, and in an oscillation
at kHz frequencies there is not enough time for any response other
than local movement of the boundary in the radial direction. Therefore particle transformation is required in far-separated areas despite the 
approximate global conservation of the amount of each
of the two forms of matter. The simple 
cylinder and piston model of Sec.~\ref{sec:piston} is a valid approximation
for a small volume element that straddles the interface between the
two phases. To use the results from Sec.~\ref{sec:piston}
we simply need to use the appropriate expression for
the local gravitational acceleration $g$.
The general relativistic generalization of the Newtonian 
hydrostatic equation is the Oppenheimer-Volkoff (OV) equation \cite{Oppenheimer:1939ne}
\beq
\ba{rcl}
\dsp\frac{d p}{d r} &=& - g_{\rm eff}(r)\, \eps(r) \\[2ex]
g_{\rm eff}(r) &=& \dsp \frac{GM}{r^2} \left( 1+\frac{p}{\epsilon} \right) 
 \left( 1+\frac{4\pi p r^3}{M} \right) 
 \left( 1-\frac{2 GM}{r}\right)^{\!-1}
\ea
\label{eqn:OV}
\eeq
where $p$, $\epsilon$ and $M$ are given by $(dM)/(dr)=4\pi r^2 \epsilon$, depending on the radial position. The effective gravitational acceleration $g_{\rm eff}$ contains general relativistic corrections to its Newtonian value $GM/r^2$.

\subsection{Movement of the ideal boundary}

We now calculate the dissipation of the energy of an r-mode in a star
with a high-density core surrounded by a low-density mantle. (In the next section
we look at the case where the phases are quark matter and hadronic matter.)
We are interested in situations where phase conversion dissipation
becomes important in r-mode oscillations when their amplitude is still
fairly low (we see in Sec.~\ref{sec:v_discussion} that this may
indeed happen), so we assume
$\delta p \ll \bar{p}$ in the region near the boundary. 
Therefore, we only need the EoS in a narrow pressure range
around the critical pressure. The EoS
can be expanded to linear order analogous to Eq.~(\ref{eqn:EoShyb})
so the pressure oscillation is given by
\beq
\delta p=\frac{\bar{\ep}}{A}\frac{\delta \ep}{\bar \ep}.
\label{eqn:dp_0_rmode}
\eeq
When $\delta \ep>0$, according to Eqs.~\eqn{eqn:de_e_rmode}--\eqn{eqn:dp_0_rmode} the r-mode
pressure oscillation in the low-density phase is 
\beq
\delta p_{\rm L}(r, \theta, \phi, t)=\bar{\ep}_{\rm L}(r)C(r)\alpha\sin^2\!\theta\cos \theta\cos (2\phi+\omega t),
\label{eqn:dp_1_rmode}
\eeq
where 
\beq
C(r)\equiv\sqrt{\frac{105}{756\pi}}\Omega ^{2} \frac{r^3}{R}.
\label{eqn:C_rmode}
\eeq

The ideal (i.e.,~long-run equilibrium at given pressure) position of the boundary $R_{\rm ib}$, analogous to $x_{\rm ib}$ in Sec.~\ref{sec:piston},
is determined by the r-mode pressure oscillation in the low-density phase, and therefore depends on the angular co-ordinates. If we write $R_{\rm ib}=\bar R_{\rm b} + \delta R_{\rm ib}$, where $\bar R_{\rm b}$ is the equilibrium position of the phase boundary with no pressure oscillation, then
from Eqs.~\eqn{eqn:OV} and \eqn{eqn:dp_1_rmode}
\beq
\delta R_{\rm ib}(t) = 
\frac{\de p_{\rm L}}{dp/dr(\bar R_{\rm b})}
= \frac{\al C_{\rm b}}{g_{\rm b}}
  \sin^2\!\theta\cos \theta\cos (2\phi+\omega t),
\label{eq:delta-R}
\eeq
where
\beq
\ba{rcl}
g_{\rm b} &\equiv& g_{\rm eff}(\bar R_{\rm b}), \\[2ex]
C_{\rm b} &\equiv& \dsp C(\bar R_{\rm b})
 =\sqrt{\frac{105}{756\pi}}\Omega ^{2} \frac{\bar R_{\rm b}^3}{R},
\ea
\label{eqn:c_bar_rmode}
\eeq
and  $g_{\rm eff}(\bar R_{\rm b})$ is the effective gravitational acceleration at $\bar R_{\rm b}$ evaluated in the low-density phase.

The oscillation amplitude of the ideal boundary position, as a
function of latitude $\theta$ in the star, is
\beq
|\delta R_{\rm ib}| = \frac{C_{\rm b}\alpha}{g_{\rm b}}|\sin^2\!\theta\cos \theta|,
\label{eqn:dr_3_rmode}
\eeq
and the maximum value of the velocity of the ideal boundary $v_{\rm ib}^{\rm max}$ is 
\beq
v_{\rm ib}^{\rm max}=\frac{C_{\rm b}\alpha\omega}{g_{\rm b}}|\sin^2\!\theta\cos \theta|.
\label{eqn:v_max}
\eeq

\subsection{R-mode energy dissipation}
\label{sec:r-mode-diss}

We now calculate $\mathrm{d}W(\theta, \phi)$,
the energy dissipated during one oscillation cycle
in a radially oriented cylinder straddling the phase boundary, with an infinitesimal base area located at a given spherical angle. Integrating this result over solid angle will give the total
dissipation of the r-mode. We use Eqs.~(\ref{eq:W1_W2}) and (\ref{eq:delta-R}),
\beq
 \mathrm{d}W(\theta, \phi)=\mathrm {d}S\left(\frac{n_{\rm H}}{n_{\rm L}}-1\right)\Delta p_{\rm L} \int_0^\tau\!\!\cos(2\phi+\omega t)\frac{\mathrm{d}\delta R_{\rm b}(t)}{\mathrm{d}t} \mathrm{d}t,
\label{eqn:dW_0}
\eeq
where $\Delta p_{\rm L}=g_{\rm b}\ep_{\rm crit}^{\rm L}|\delta R_{\rm ib}|$, and from Eq.~\eqn{eqn:dr_3_rmode}
\beq
\Delta p_{\rm L}=\ep_{\rm crit}^{\rm L}C_{\rm b}\alpha|\sin^2\!\theta\cos \theta| \, 
\label{eqn:dpL_rmode}
\eeq
and $\mathrm {d}S=\bar{R}_{\rm b}^2\sin\theta\mathrm {d}\theta\mathrm {d}\phi$.
As discussed earlier, these estimates are based on an approximate r-mode profile. To estimate the uncertainty due to this simplification we compare two idealized cases: where the pressure oscillation is harmonic in the low-density phase and where it is harmonic in the high-density phase.
As discussed below, Eq.~(\ref{eq:W1_W2}) the difference for an infinitesimal volume element is of order $\Delta W/W\simeq\Delta \ep/\ep_{\rm L}$, which directly gives an estimate for the uncertainty of the dissipation in the case of global r-modes. Typical density steps at first-order transitions in a compact star are less than a factor of two, but due to the simplified model assumptions we make here our results should be viewed as order of magnitude estimates.

\section{Hadron-quark conversion in a hybrid star}
\label{sec:front_speed}

The damping mechanism that we have analyzed above is generic and will operate in any situation where there are two phases with a sharp interface. However, the amount of damping depends crucially on how the
interface between the two phases moves via conversion of one phase into the other. To explore a realistic case, we now estimate the boundary velocity for 
an interface between strange quark matter and nuclear matter in a hybrid star, and obtain an estimate of the resultant r-mode saturation amplitude in this scenario. 

It is worth mentioning that the scenario depicted here
of a smooth and steadily moving phase boundary might be disturbed by instabilities that lead to a turbulent burning front. In the conversion of a neutron star to a quark star, this instability occurs and may lead to a detonation \cite{Horvath:1988nb} or a deflagration \cite{Drago:2005yj,Herzog:2011sn,Pagliara:2013tza}. 
That full conversion process occurs when quark matter is the preferred
state all the way down to zero pressure (the ``strange matter hypothesis'').
In our situation, however, the conversion is much slower: We study the case where conversion of hadronic matter to quark matter occurs only at a high critical pressure. The movement of the phase boundary in our case 
is driven by a small deviation from equilibrium, induced by oscillations.
We defer the study of turbulent instabilities in this context
to future work.

We use the calculational techniques developed by Olinto
\cite{Olinto:1986je} to study the movement of the phase boundary in the
strange matter hypothesis scenario, but we are interested in conversion of
quark matter to nuclear matter as well as nuclear matter to quark matter,
since both processes occur as our burning front moves inwards and outwards
periodically in response to an oscillation in the pressure.

\subsection{Pressure and chemical potential at the interface}
\label{sec:p_mu_relation}

\begin{figure}[htb]
\includegraphics[width=\hsize]{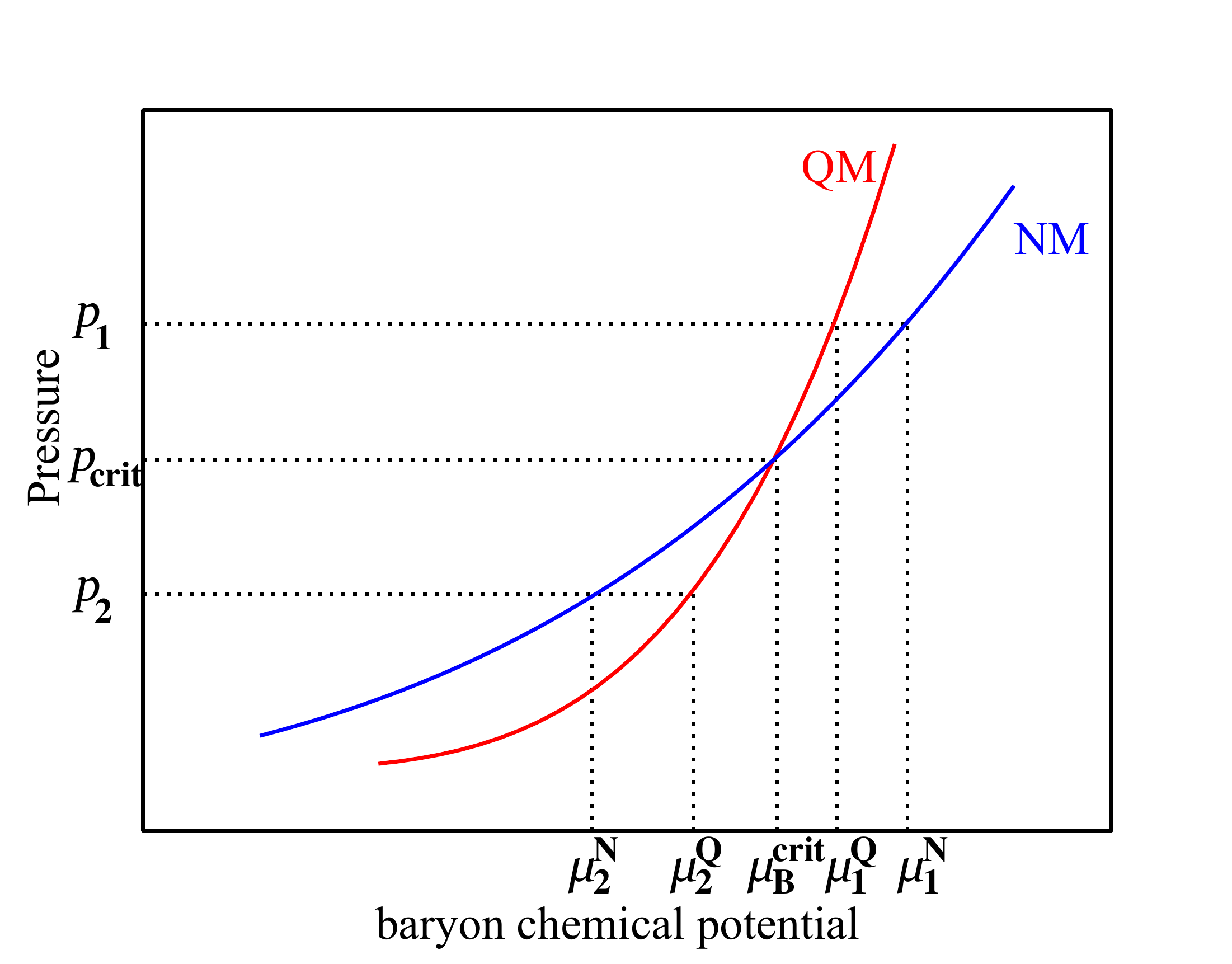}
\caption{(Color online) Schematic plot of the pressure as a function of baryon chemical potential in $\beta$-equilibrated ($\mu_{\rm K}=0$) nuclear matter and quark matter. At a given pressure, the
phase with lower $\mu_{\rm B}$ is thermodynamically favored.}
\label{fig:nm_qm_transition}
\end{figure}

In equilibrium, both pressure and baryon chemical potential are continuous
across the phase boundary between nuclear and quark matter, and
their values at the boundary are the critical values at which the phase transition
occurs ($p=p_{\rm crit}, \mu_{\rm B}=\mu_{\rm B}^{\rm crit}$). When the system
is driven out of equilibrium by global pressure oscillations, the 
phase boundary
may temporarily be at a different pressure because the conversion
between nuclear and quark matter has a limited rate. The boundary is then
out of chemical equilibrium, and the baryon chemical potential is no longer
continuous at the boundary because baryon number cannot
flow freely through the boundary. On the time scale of
chemical equilibration the pressure is still
continuous because it equilibrates at the speed of sound, which is
of order $c$. The burning front will move 
as the phase with higher baryon chemical potential converts into the
phase with lower baryon chemical potential.
The situation is illustrated in Fig.~\ref{fig:nm_qm_transition}.
If the pressure at the boundary is
above the critical value ($p_{\rm b}=p_{1}>p_{\rm crit}$), the baryon number
chemical potential in quark matter is lower ($\mu_{1}^{\rm Q}<\mu_{1}^{\rm
  N}$). Nuclear matter (NM) is 
then converted into quark matter
(QM) and the front moves outwards. If the pressure at the boundary
is below the critical value ($p_{\rm b}=p_{2}<p_{\rm crit}$) the front moves in
the opposite direction converting quark matter back into nuclear matter.

As we see below,
when the boundary is out of chemical equilibrium, and moving to reestablish
that equilibrium, it has around it a
$\rm {NM}\rightleftarrows \rm {QM}$ conversion region, where 
the matter is out of $\beta$ equilibrium.

The chemical equilibration of quark matter can proceed via the
nonleptonic channel $u+s\leftrightarrow d+u$ and also the leptonic Urca
channel $d \to u+e^{-}+\bar\nu$ and $u+e^{-} \to d+\nu$.
Following Ref.~\cite{Olinto:1986je}, we neglect the
Urca channel here for simplicity, but we discuss its potential impact in 
Sec.~\ref{sec:conclusion}.
Nonleptonic $\beta$-equilibration processes are driven by
the chemical potential $\mu_{\rm K}$, which couples to the imbalance
between strange and down quarks; $\mu_{\rm K}$ is zero in $\beta$-equilibrated matter,
but not in the conversion region,
\beq
\ba{rcl}
\mu_{\rm K}&\equiv&\mu_{\rm d}-\mu_{\rm s} \ ,\\
n_{\rm K} &=& \half(n_{\rm d}-n_{\rm s}) \ .
\ea
\label{eqn:muK}
\eeq
In the following sections we discuss how $\mu_{\rm K}$ and $\mu_{\rm B}$ vary in the conversion region when the front is moving, in order to estimate the speed of the boundary in two half cycles of oscillation, which determines the energy dissipation over the complete period.

\begin{figure*}
\parbox{0.5\hsize}{
\vspace{-2ex}
\includegraphics[width=\hsize]{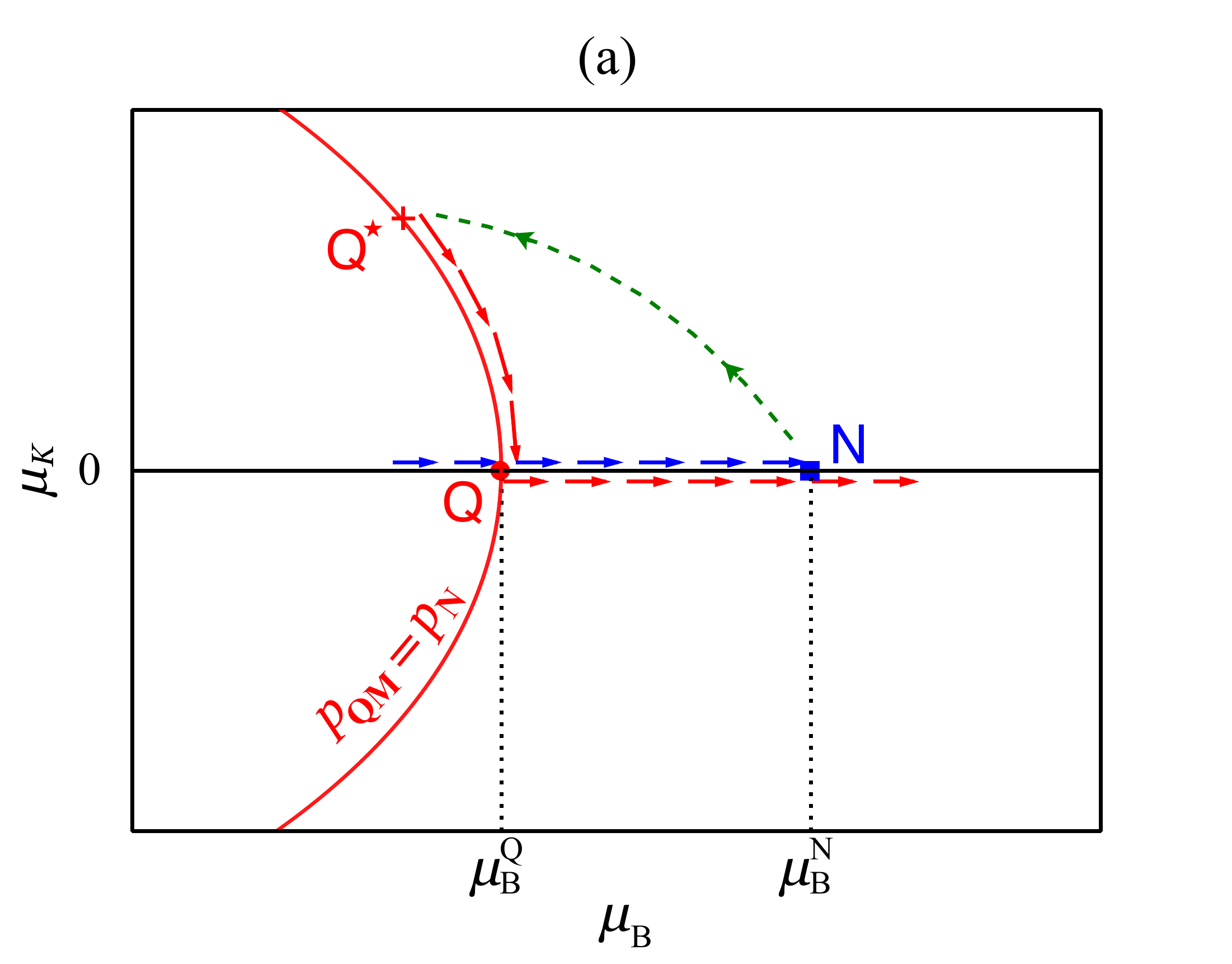}\\[2ex]
}\parbox{0.5\hsize}{
\vspace{-2ex}
\includegraphics[width=\hsize]{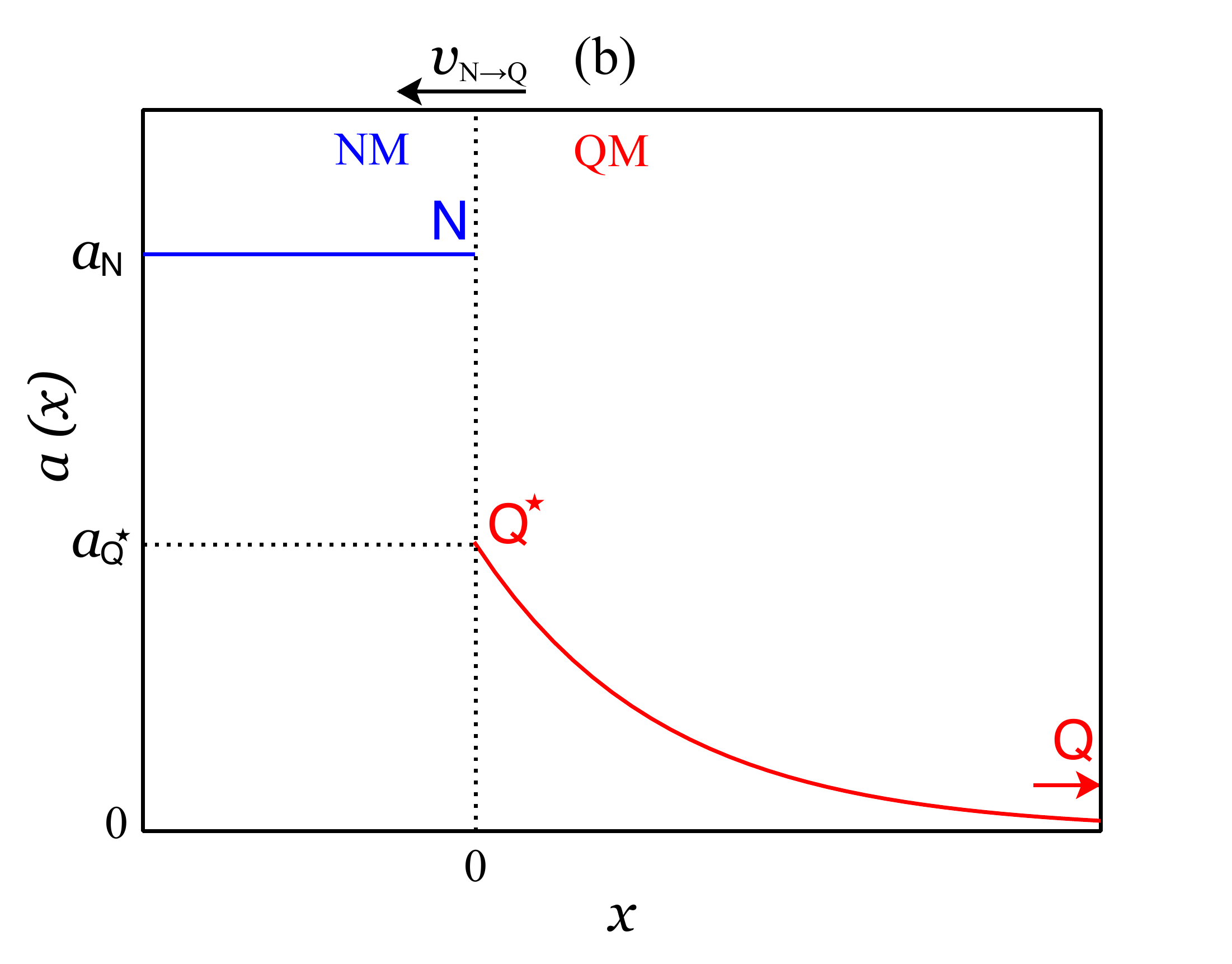}\\[2ex]
}
\caption{(Color online) Conversion of nuclear matter into quark matter.
(b): Spatial variation of the $K$-fraction parameter $a$ [Eq.~\eqn{eqn:a_x}]
in the conversion region where the pressure is above $p_{\rm crit}$ (see Fig.~\ref{fig:nm_qm_transition}). (a): Corresponding
path in the $(\mu_{\rm B},\mu_{\rm K})$ plane of chemical potentials. The quark matter isobar (red curve passing through $\rm Q$ and $\rm Q^*$) is at the same pressure as the equilibrated nuclear matter (point $\rm N$), and the arrows follow increasing pressure except
from N to $\rm Q^*$ to $\rm Q$ where pressure is constant (traversing
increasing $x$ in the right panel). 
}
\label{fig:nm-qm}
\end{figure*} 

\begin{figure*}
\parbox{0.5\hsize}{
\vspace{-2ex}
\includegraphics[width=\hsize]{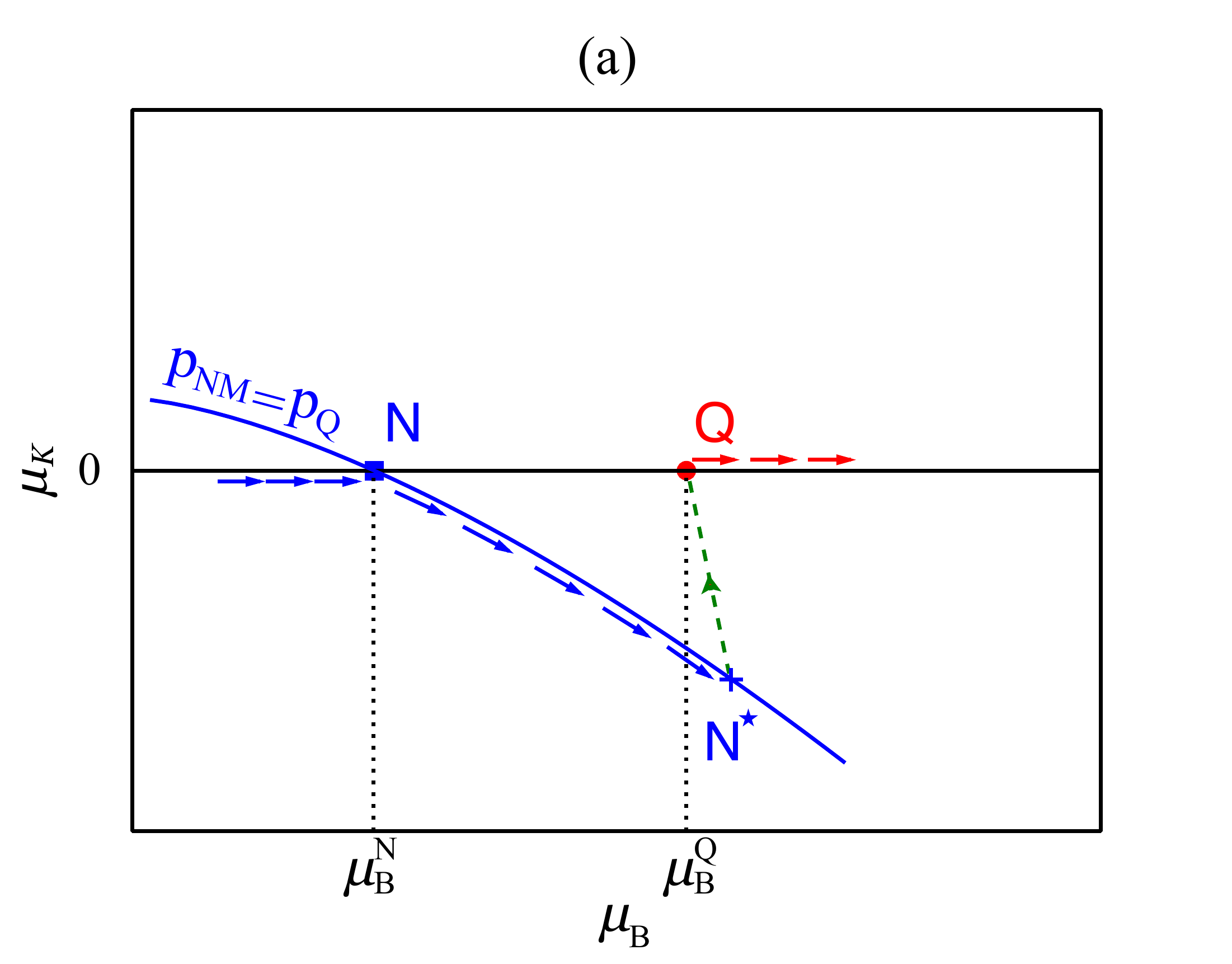}\\[2ex]
}\parbox{0.5\hsize}{
\vspace{-2ex}
\includegraphics[width=\hsize]{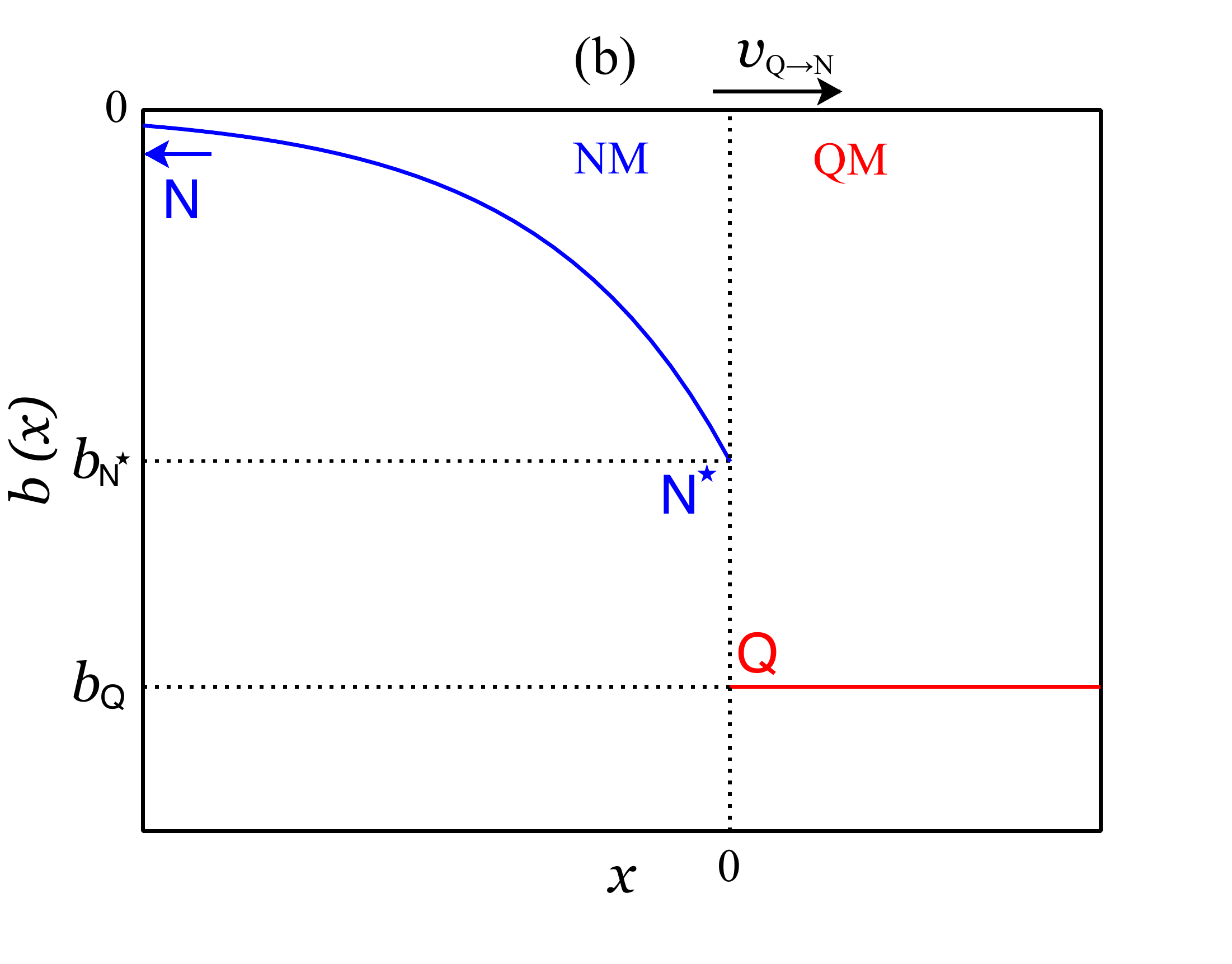}\\[2ex]
}
\caption{(Color online) Conversion of quark matter into nuclear matter.
(b): Spatial variation of 
the $K$-fraction parameter $b$ [Eq.~\eqn{eqn:b_x}]
in the conversion region where the pressure is below $p_{\rm crit}$ (see Fig.~\ref{fig:nm_qm_transition}). (a): Corresponding
path in the $(\mu_{\rm B},\mu_{\rm K})$ plane of chemical potentials. The nuclear matter isobar (blue curve passing through $\rm N$ and $\rm N^*$) is at the same pressure as the equilibrated quark matter (point $\rm Q$), and the arrows follow increasing pressure except from N to $\rm N^*$ to $\rm Q$ where pressure is constant (traversing
increasing $x$ in the right panel).
}
\label{fig:qm-nm}
\end{figure*}

\subsection{Conversion of nuclear matter into quark matter}
\label{sec:nm_to_qm}

To estimate the front speed in the $\rm NM\to \rm QM$ transition when $p_{\rm
  b}=p_{1}>p_{\rm crit}$, we use the one-dimensional steady-state
approximation used by Olinto \cite{Olinto:1986je} to study the
irreversible conversion of nuclear matter to strange matter in the
scenario where strange quark matter is stable at zero pressure.
In our scenario there is a critical pressure at which the transition
occurs, with conversion going in either direction depending on the
history of the system, but Olinto's approach is still applicable.
The analysis is
conveniently performed in the rest frame of the boundary, where the boundary is at
$x=0$, neutron matter is at $x<0$ and strange quark matter at $x>0$. 
The transformation of neutron matter into strange quark matter requires
considerable strangeness production, which can only be accomplished by
flavor-changing weak interactions. The slow rate of weak interactions means that at the
front nuclear matter is converted in to some form of non-$\beta$-equilibrated
quark matter (with $\mu_{\rm K}\neq 0$).
In the conversion region behind the front there are flavor-changing non-leptonic
interactions and strangeness diffusion.
The weak interactions create strangeness and
allow $\mu_{\rm K}$ to return to zero over
a distance scale of order $(D_{\rm Q}\tau_{\rm Q})^{1/2}$ where $\tau_{\rm Q}$ is the
time scale of the flavor-changing nonleptonic
interactions and $D_{\rm Q}$ is the
diffusion constant for flavor. The diffusion of strangeness
towards the boundary and "downness" away from the boundary
allows the strange matter at the boundary to have
a strangeness fraction different from that of the nuclear matter,
which is undergoing deconfinement as the front moves.

In general, strangeness gradients could also exist in front of the boundary,
as strangeness could diffuse through the boundary, creating (or adding to)
hyperons on the nuclear matter side. However, 
following Ref.~\cite{Olinto:1986je}, we assume that the front
moves fast enough for this effect to be negligible, so 
$\mu_{\rm K}=0$ everywhere ahead of the moving front, i.e.,~at $x<0$.
The conversion region is then limited to $x>0$ and 
can be characterized by $\mu_{\rm K}(x)$, or
equivalently by the $K$-fraction
parameter $a(x)$, which decreases with increasing strangeness fraction
\beq
a(x)\equiv \frac{n_{\rm K}(x)-n_{\rm K}^{\rm Q}}{n_{\rm Q}},
\label{eqn:a_x}
\eeq
where $n_{\rm Q}$ is the baryon number density in equilibrated strange 
quark matter, and the $K$ density is $n_{\rm K}^{\rm Q^{\star}}$ at $x=0$ and 
as $x\to \infty$ it grows asymptotically to the constant value 
$n_{\rm K}^{\rm Q}$ for equilibrated strange 
quark matter. From now on for simplicity we always assume that there are equal numbers of up, down, and strange quarks in equilibrated quark matter ($n_{\rm K}^{\rm Q}=0$). In equilibrated nuclear matter there are only up and down quarks (we assumed no hyperons in front of the boundary), so $n_{\rm K}=n_{\rm d}/2=n_{\rm N}$ for $x<0$.

The spatial variation of $a$ (Fig.~\ref{fig:nm-qm}, right panel)
is determined by the steady-state
transport equation, written in the rest frame of the boundary,
\bea
D_{\rm Q}\,a''-v_{\rm{N\!\to\!Q}}\,a'-\mathcal{R}_{\rm Q}(a) = 0, \nonumber \\[1ex] 
\mathcal{R}_{\rm Q}(a) = (\Gamma_{{\rm d}\to {\rm s}}-\Gamma_{{\rm s}\to {\rm d}})/n_{\rm   Q} ,
\label{eqn:Deq_a}
\eea
where $D_{\rm Q}$ is the flavor diffusion coefficient, $v_{\rm{N\!\to\!Q}}$ is
the front speed, and $\mathcal{R}_{\rm Q}(a)$ is the 
net rate of flavor-changing weak interactions.
The boundary conditions are 
\beq
\ba{rcl@{\qquad}l}
a(0^{-})&=&\displaystyle \frac{n_{\rm N}}{n_{\rm Q}}\equiv a_{\rm N},\!\!\!\!\!
  & a(x\to \infty)\to 0, \\[2ex]
a(0^{+})&=&\displaystyle \frac{n_{\rm K}^{\rm Q^{\star}}}{n_{\rm Q}}\equiv a_{\rm Q^{\star}},
 & a'(0^+)=-v_{\rm{N\!\to\!Q}}\displaystyle \left(\frac{a_{\rm N}-a_{\rm Q^{\star}}}{D_{\rm Q}}\right). \ 
\ea
\label{eqn:NtoQ-BC}
\eeq

To understand the discontinuity in $a(x)$ across the boundary, let us consider how
the chemical potentials vary in the conversion region. 
The left panel of
Fig.~\ref{fig:nm-qm} shows a schematic plot in the 
$(\mu_{\rm B}, \mu_{\rm K})$ plane.
The parabolic-looking curve is the quark matter isobar
for pressure $p_{\rm b}>p_{\rm  crit}$. 
The square marked ``N'' is $\beta$-equilibrated ($\mu_{\rm K}=0$)
nuclear matter at the same pressure.
The spatial variation in the conversion region, shown in the right
panel of Fig.~\ref{fig:nm-qm}, can then be mapped on the chemical potential
space as follows. At $x<0$ we have $\beta$-equilibrated nuclear matter
(N). At $x=0$, where $a(x)$ drops from $a_{\rm N}$ to $a_{\rm Q^{\star}}$, 
$\mu_{\rm K}$ jumps
to $\rm Q^{\star}$, which is  out-of-equilibrium quark matter with nonzero
$\mu_{\rm K}$, but at the same pressure as the nuclear matter. Then as we
traverse the conversion region (increasing $x$), $\mu_{\rm K}$ decays
to zero, finally arriving at equilibrated quark matter (Q).
All of these configurations are at the same
pressure, based on the assumption that the thickness of the conversion region is
negligible when compared to the radius of the star. 
The arrows along the $\mu_{\rm K}=0$ axis show how $\mu_{\rm B}$ varies as
one moves larger distances through $\beta$-equilibrated matter on either
side of the conversion region, with the pressure rising monotonically.
The arrows above the $\mu_{\rm K}=0$ axis (blue online)
show $\mu_{\rm B}$ increasing as we move inwards through nuclear matter until at N
($\mu_{\rm B}=\mu_{\rm B}^{\rm N}$)
we reach the phase boundary. After traversing the phase boundary 
and conversion region as described above,
we are at Q, in $\beta$-equilibrated quark matter at lower $\mu_{\rm B}$, and as we
move into the quark core, $\mu_{\rm B}$ rises again (arrows below $\mu_{\rm K}=0$ axis,
red online).

Olinto \cite{Olinto:1986je} argued that when the phase boundary is 
in a steady state of motion there is a ``pileup'' of nuclear matter
in front so that nuclear and quark matter have the same density there,
and the boundary has the same velocity relative to
nuclear matter and quark matter; i.e., the nuclear matter near the
boundary is stationary
relative to the quark matter. However, we argue that this is not possible
in steady state. When the phase boundary moves, part of the star is transformed from lower density nuclear matter to denser quark matter, and hydrostatic equilibrium requires the star to shrink. This means that in the outer
parts of the star the nuclear matter must
fall inwards under gravity, so it is moving towards the quark matter. If the
inward velocity of the nuclear matter went to zero near the phase boundary,
this would require that the ``pileup'' grows with time, which is not
a steady state situation.
Instead, we argue that the baryon number conservation condition is automatically fulfilled because the weight of the outer region of the star pushes nuclear matter in to the front as fast
as the front can ``consume'' it. The density of nuclear matter at the
boundary is therefore unchanged by the movement of the boundary, and the
nuclear matter velocity takes the value that is determined by baryon number
conservation. As we saw in Eq.~(\ref{eq:W1_W2}) this density step at the phase boundary is crucial 
for phase-conversion dissipation to occur ($a_{\rm N}<1$).

For a fixed value of $a_{\rm Q^{\star}}$, there is
only one $v_{\rm{N\!\to\!Q}}$ which guarantees a solution to
Eq.~\eqn{eqn:Deq_a} that satisfies the boundary conditions. To find the proper $v_{\rm{N\!\to\!Q}}$, we apply the method in \cite{Olinto:1986je}, which analogizes Eq.~\eqn{eqn:Deq_a} to a classical mechanical problem and solves for the correct potential term, transforming the boundary value problem into an initial value problem. Taking into account both subthermal ($\mu_{\rm K}\ll T$) and suprathermal ($\mu_{\rm K}\gg T$) regimes in the weak rate \cite{Alford:2010gw}, the analytical approximation for the front speed in the $\rm {NM\to QM}$ half cycle is
\beq
v_{\rm{N\!\to\!Q}}\simeq\sqrt{\frac{D_{\rm Q}}{\tau_{\rm Q}}\frac{a_{\rm Q^{\star}}^4+2\eta
_{\rm Q} a_{\rm Q^{\star}}^2}{2a_{\rm N}(a_{\rm N}-a_{\rm Q^{\star}})}},
\label{eqn:v_qm_original}
\eeq
where $D_{\rm Q}$ is the diffusion constant for flavor, $\tau_{\rm Q}$ is the
time scale of nonleptonic flavor-changing interactions, and $\eta_{\rm Q}$ gives the ratio of subthermal to suprathermal rates. Equation~\eqn{eqn:v_qm_original} is a generalization of Eq.~(12) in Ref.~\cite{Olinto:1986je}, which is only valid in the suprathermal regime. As we see later on, $a_{\rm Q^{\star}}$ is much less than $a_{\rm N}$; therefore Eq.~\eqn{eqn:v_qm_original} becomes
\beq
v_{\rm{N\!\to\!Q}}\simeq\frac{1}{a_{\rm N}}\sqrt{\frac{D_{\rm Q}}{2\tau_{\rm Q}}}\sqrt{a_{\rm Q^{\star}}^4 +2\eta_{\rm Q} a_{\rm Q^{\star}}^2}.
\label{eqn:v_qm}
\eeq
The full rate for the nonleptonic strangeness-changing
process has been computed in Ref.~\cite{Madsen:1993xx}, yielding
\bea 
\mathcal{R}_{\rm Q}(a) &\simeq& (a^3+\eta_{\rm Q} a)/\tau_{\rm Q},\ \\
\eta_{\rm Q}&=&\frac{9\pi^2 T^2}{\mu_{\rm Q}^2},\ \\
\tau_{\rm Q} &=&\left(\frac{128}{27\times5 \pi^3} G_{F}^{2}\cos^{2}\!\theta_{c}\sin^{2}\!\theta_{c}\mu_{\rm Q}^{5}\right)^{-1},
\label{eqn:tau_qm}
\eea
where $G_F$ is the Fermi constant, $\theta_c$ is the Cabibbo angle, and therefore $\tau_{\rm Q}\simeq1.3\times10^{-9}\,\rm s\, \left(300\,{\rm MeV}/\mu_{\rm Q}\right)^{5}$, to leading order the diffusion coefficient (see Eqs. (28) and (36) in Ref.~\cite{Heiselberg:1993cr}) 
\beq
D_{\rm Q}\simeq \frac{\pi q_{D}^{2/3}}{24\;\alpha_{s}^2 \;h\; T^{5/3}}
\label{eqn:D_qm}
\eeq
where $h=\Gamma(\frac{8}{3})\zeta(\frac{5}{3})(2\pi)^{2/3}\simeq 1.81$, $\alpha_{s}=g^2/4\pi$ is the QCD coupling constant, and the Debye wave number for cold quark matter of three flavors is $q_D$ where $q_{D}^2=3g^2\mu^2/(2\pi^2)$. The temperature dependence $T^{-5/3}$ comes from Landau damping that dominates for $T\ll \mu$ compared to the Debye screened case $D\propto T^{-2}$.

Different values of $a_{\rm Q^{\star}}$ give
different front profiles corresponding to
different front velocities. There is an upper limit on $a_{\rm Q^{\star}}$ which is constrained by the amplitude of external pressure oscillation, and the argument is as follows:

In order for the boundary to move, it must be favorable for neutrons to
turn in to quarks at the boundary, so the total chemical potential per unit
 baryon number must be larger in $\beta$-equilibrated nuclear matter (N in Fig.~\ref{fig:nm-qm}) than in out-of-equilibrium quark matter ($\rm Q^{\star}$) \footnote{We thank Mikhail Gusakov for emphasizing this point via personal communications.}
\beq
\mu_{\rm B}^{\rm N}> \mu_{\rm B}^{\rm {Q^{\star}}} + \frac{n_{\rm K}^{\rm {Q^{\star}}}}
{n_{\rm Q^{\star}}}\mu_{\rm K}^{\rm {Q^{\star}}}.
\label{eqn:chem_potl_a0}
\eeq
On the isobar for quark matter, we parametrize the pressure at ($\mu_{\rm B}$, $\mu_{\rm K}$) as an expansion near equilibrium ($\rm Q$)
\begin{align}
p_{\rm QM}(\mu_{\rm B}, \mu_{\rm K})&=p_{\rm Q}+n_{\rm Q}(\mu_{\rm B}-\mu_{\rm B}^{\rm Q})+n_{\rm K}^{\rm Q}(\mu_{\rm K}-\mu_{\rm K}^{\rm Q}) \nonumber \\
&+\frac{1}{2}\chi_{\rm K}^{\rm Q}(\mu_{\rm K}-\mu_{\rm K}^{\rm Q})^2+\cdots
\label{eqn:p1_qm}
\end{align}
where $\chi_{\rm K}^{\rm Q}\equiv\partial n_{\rm K}/\partial \mu_{\rm K}$ is the susceptibility with respect to $K$-ness evaluated at equilibrium ($\rm Q$). 
In equilibrated quark matter, $\mu_{\rm K}^{\rm Q}=0$. Since the whole
conversion region is at the same pressure $p_{\rm Q^{\star}}=p_{\rm Q}=p_{\rm b}$, solving for $\mu_{\rm B}^{\rm Q^{\star}}$ we have 
\beq
\mu_{\rm B}^{\rm Q^{\star}}=\mu_{\rm B}^{\rm Q}-\left(\frac{n_{\rm K}^{\rm Q}}{n_{\rm Q}}-\frac{\chi_{\rm K}^{\rm Q}\mu_{\rm K}^{\rm Q^{\star}}}{2\,n_{\rm Q}}\right) \mu_{\rm K}^{\rm Q^{\star}}.
\label{eqn:muB_Q_star}
\eeq
Assuming that $\rm Q^{\star}$ is close to equilibrium, so that $n_{\rm K}^{\rm Q^{\star}}\approx n_{\rm K}^{\rm Q}$, $n_{\rm Q^{\star}}\approx n_{\rm Q}$, Eq.~\eqn{eqn:chem_potl_a0} becomes
\beq
\mu_{\rm B}^{\rm N}-\mu_{\rm B}^{\rm Q}>\frac{\chi_{\rm K}^{\rm Q} \left(\mu_{\rm K}^{\rm Q^{\star}}\right)^2}{2n_{\rm Q}} \ .
\label{eqn:dmu_a0}
\eeq
From Eq.~\eqn{eqn:NtoQ-BC}, $a_{\rm Q^{\star}}=n_{\rm K}^{\rm Q^{\star}}/n_{\rm Q}\approx\mu_{\rm K}^{\rm Q^{\star}}\chi_{\rm K}^{\rm Q}/n_{\rm Q}$, and then Eq.~\eqn{eqn:dmu_a0} leads to an upper bound on $a_{\rm Q^{\star}}$
\beq
a_{\rm Q^{\star}}^{\rm max}=\sqrt{\frac{2\Delta\mu_{\rm B}\chi_{\rm K}^{\rm Q}}{n_{\rm Q}}},
\label{eqn:a0_max}
\eeq
with
\beq
\Delta\mu_{\rm B}\equiv\mu_{\rm B}^{\rm N}-\mu_{\rm B}^{\rm Q}\simeq(\gamma-1)\delta p/n_{\rm Q}, 
\label{eqn:dmu_p}
\eeq
where $\delta p=|p_{\rm b}-p_{\rm crit}|\geqslant 0$
 (Fig.~\ref{fig:nm_qm_transition}) and $\gamma\equiv n_{\rm Q}/n_{\rm N}=1/a_{\rm N}$.
Notice that the derivation of Eq.~\eqn{eqn:a0_max} is totally general and can also be applied to matter with nonzero $n_{\rm K}$ at equilibrium (see Sec.~\ref{sec:qm_to_nm}).

\begin{figure}[htb]
\includegraphics[width=\hsize]{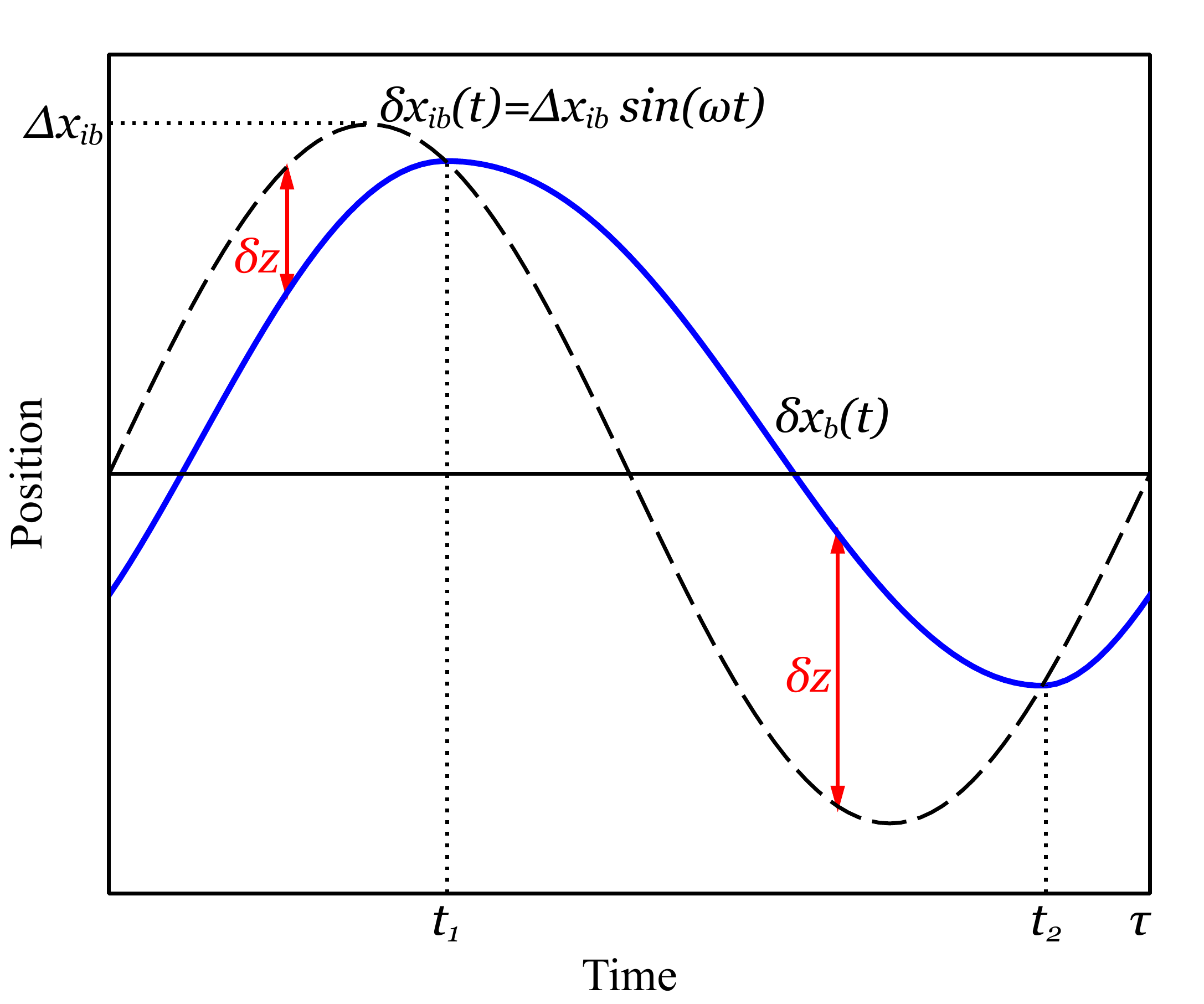}
\caption{(Color online) Diagram showing how the ideal boundary position (dashed line) and the real boundary position (solid [blue] line) vary in time. The ideal boundary is where the phase boundary would be if the phase conversion process equilibrated instantaneously, and it is determined by the instantaneous external pressure [Eq.~\eqn{eqn:ideal_wall}]. The real boundary is always ``chasing'' the ideal
boundary, with velocity given by Eq.~\eqn{eqn:dx_t_qm} and \eqn{eqn:dx_t_nm}
where $\de z(t)$ is its distance from the ideal boundary. The real boundary coincides with the ideal boundary twice per cycle, at $t=t_1$ and $t=t_2$.}
\label{fig:x_ideal_real}
\end{figure}

The pressure oscillation is related to how far the real boundary is away from its ideal position via
\beq
\delta p=g_{\rm b}\ep_{\rm crit}^{\rm N}\bigl(\delta x_{\rm ib}(t)-\delta x_{\rm b}(t)\bigr),
\label{eqn:dp_dx_qm}
\eeq
where the ideal position
$\delta x_{\rm ib}(t) = \Delta x_{\rm ib} \sin(\omega t)$, 
and the amplitude $\Delta x_{\rm ib} = \Delta p_{\rm N}/g_{\rm b}\ep_{\rm crit}^{\rm N}$ [see Eq.~\eqn{eqn:ideal_wall}]. If we assume that the boundary is always moving at its maximum speed ($a_{\rm Q^{\star}}\approx a_{\rm Q^{\star}}^{\rm max}$), then according to Eqs.~\eqn{eqn:v_qm} and \eqn{eqn:a0_max}--\eqn{eqn:dp_dx_qm}
the velocity of the boundary in the $\rm NM\to QM$ half cycle 
($t<t_1$ and $t>t_2$ in Fig.~\ref{fig:x_ideal_real}) is determined by
\beq
\frac{\mathrm {d} \delta x_{\rm b}}{\mathrm {d}t}
\simeq \frac{1}{a_{\rm N}}\sqrt{\frac{D_{\rm Q}}{2\tau_{\rm Q}}}\sqrt{\bigl[\de z / \ell_{\rm Q} \bigr]^2+2\eta_{\rm Q}\de z / \ell_{\rm Q}},
\label{eqn:dx_t_qm} 
\eeq
where $\de z \equiv |\delta x_{\rm ib}-\delta x_{\rm b}|$
is how far the boundary is from its equilibrium 
position at the current pressure (see Fig.~\ref{fig:x_ideal_real}), and $\ell_{\rm Q}$ characterizes its typical length
\beq
\ell_{\rm Q}= \frac{(n_{\rm Q}/\chi_{\rm K}^{\rm Q})n_{\rm Q}}{2(\gamma-1)g_{\rm b}\ep_{\rm crit}^{\rm N}}.
\label{eqn:a_tilde_qm}
\eeq

\subsection{Conversion of quark matter into nuclear matter}
\label{sec:qm_to_nm}

The conversion from quark matter to nuclear matter has not been
analyzed previously because it does not arise 
if strange matter is stable at zero pressure, which is the context
in which previous analyses were performed.
However, it can analogously
be described in terms of conversion and diffusion
behind the boundary, now on the hadronic side where strangeness is carried by
hyperons. There
are various hyperons that could be present in dense hadronic matter and
correspondingly multiple weak reactions involving these hyperons. For an illustrative calculation we consider one such process,
$n+n \to p+\Sigma^{-}$, which is a reasonable choice because
$\Sigma^{-}$ hyperons are expected to be among the first to appear when
nuclear matter is compressed; see, e.g.,~Ref.~\cite{Schaffner:1995th}.
For simplicity, we also neglect electrons
in the system, which is admittedly not a good approximation, but
we are only aiming to provide an illustrative example.
In this case 
\bea
\mu_{\rm K} &=& 2\mu_{\rm n}-\mu_{\rm p}-\mu_{\rm \Sigma} \\
n_{\rm K} &=&\frac{1}{6}(2n_{\rm n}-n_{\rm p}-n_{\rm \Sigma}) \ .
\eea
Moving away from the boundary on the hadronic side, into the
conversion region, the $K$ density is $n_{\rm K}^{\rm N^{\star}}$ at $x=0$ and 
as $x\to -\infty$ it grows asymptotically to the nonzero constant value 
$n_{\rm K}^{\rm N}\simeq n_{\rm n}/3\approx n_{\rm N}/3$ for equilibrated nuclear matter with $\Sigma^{-}$ hyperons, where $n_{\rm N}$ is the baryon number density in equilibrated nuclear matter. As before, we neglect strangeness conversion ahead of the boundary, which is the quark matter region in this case. In equilibrated quark matter we assume there are equal numbers of up, down, and strange quarks, so $n_{\rm K}=0$ for $x>0$.

As in Sec.~\ref{sec:nm_to_qm} we define
a parameter to characterize the 
deviation of the $K$ fraction from its equilibrium value,
\beq
b(x)\equiv\frac{n_{\rm K}(x)-n_{\rm K}^{\rm N}}{n_{\rm N}},
\label{eqn:b_x}
\eeq
and the steady-state transport equation for $b(x)$ in the rest frame of the boundary is
\bea
D_{\rm N}\,b''-v_{\rm{Q\!\to\!N}}\,b'-\mathcal{R}_{\rm N}(b)=0 ,
\label{eqn:Deq_b}  \\[1ex]
\mathcal{R}_{\rm N}(b)=(\Gamma_{\rm {n+n \to p+\Sigma^{-}}}-\Gamma_{\rm {p+\Sigma^{-}\!\to n+n}})/n_{\rm N},
\eea
where $D_{\rm N}$ is the flavor diffusion coefficient, 
$v_{\rm{Q\!\to\!N}}$ is the
front speed for the ${\rm QM}\to{\rm NM}$ transition, and $R_{\rm N}(b)$ the 
strangeness-changing
reaction rate divided by the baryon number density in nuclear matter.
The boundary conditions are
\beq
\ba{rcl@{\qquad}l}
b(0^{-})&=&\displaystyle \frac{n_{\rm K}^{\rm N^{\star}}-n_{\rm K}^{\rm N}}{n_{\rm N}}\equiv b_{\rm N^{\star}},\!\!\!\!\!
  & b(x\to -\infty)\to 0, \\[2ex]
b(0^{+})&=&\displaystyle \frac{-n_{\rm K}^{\rm N}}{n_{\rm N}}\equiv b_{\rm Q},
 & b'(0^-)=v_{\rm{Q\!\to\!N}}\displaystyle \left(\frac{b_{\rm Q}-b_{\rm N^{\star}}}{D_{\rm N}}\right). \
\ea
\label{eqn:QtoN-BC}
\eeq

The right panel of Fig.~\ref{fig:qm-nm} shows how $b(x)$ varies through
the phase boundary and transition region.
The left panel of Fig.~\ref{fig:qm-nm} shows schematically the 
behavior in the $(\mu_{\rm B}, \mu_{\rm K})$ plane.
The short curve through N and N${}^*$
is the nuclear matter isobar for pressure
$p_{\rm b}<p_{\rm crit}$. The dot marked ``Q'' is 
$\beta$-equilibrated quark matter at the same pressure,
which exists (see right panel) at $x=0^{+}$.
The point N${}^{\star}$ is out-of-equilibrium nuclear matter, 
which is found just behind the boundary at $x=0^{-}$.
The point N is $\beta$-equilibrated nuclear matter, which
is found at the tailing end of the conversion region. 
All these forms of matter are at the same pressure as long as thickness of the conversion region is much smaller than the radius of the star. 
The arrows represent how the chemical composition changes as one moves from 
the hadronic outer part of the star through the conversion region to the quark core. At the boundary $\mu_{\rm B}^{\rm N}<\mu_{\rm B}^{\rm Q}$ and $\mu_{\rm K}$ in out-of-equilibrium nuclear matter is negative because of the presence of massive hyperons.

Following the same logic as in the previous section, we find the analytic approximation for the velocity of the boundary
\bea
v_{\rm{Q\!\to\!N}}&\simeq&-\sqrt{\frac{D_{\rm N}}{\tau_{\rm N}}\frac{b_{\rm N^{\star}}^4+2\eta_{\rm N} b_{\rm N^{\star}}^2}{2b_{\rm Q}(b_{\rm Q}-b_{\rm N^{\star}})}}\nonumber\\
&\xrightarrow{|b_{\rm N^{\star}}|\ll |b_{\rm Q}|}&\frac{1}{b_{\rm Q}}\sqrt{\frac{D_{\rm N}}{2\tau_{\rm N}}}\sqrt{b_{\rm N^{\star}}^4 +2\eta_{\rm N} b_{\rm N^{\star}}^2},
\label{eqn:v_nm}
\eea
where $-b_{\rm Q}\lesssim1/3$. The full rate for the weak interaction has been computed in Ref.~\cite{Haensel:2001em} as
\bea
\mathcal{R}_{\rm N}(b) &\simeq&(b^3+\eta_{\rm N} b)/\tau_{\rm N},\ \\
\eta_{\rm N}&=&\frac{4\pi^2 T^2 \left(\chi_{\rm K}^{\rm N}\right)^2}{n_{\rm N}^2},
\label{eqn:eta_nm}
\eea
and the time scale
\beq
\tau_{\rm N}=\left[\frac{-2\,\chi G_{F}^{2}}{3 (2\pi)^5}\,\cos^{2}\!\theta_{c}\,\sin^{2}\!\theta_{c}\,m_{\rm n}^{\ast 2}m_{\rm p}^{\ast}\,m_{\rm \Sigma}^{\ast}\,k_{\rm F}^{\Sigma}\,n_{\rm N}^2 \left(\chi_{\rm K}^{\rm N}\right)^{-3}\right]^{-1},
\label{eqn:tau_nm}
\eeq
where $\chi$ is determined by the reduced symmetric and antisymmetric coupling constants with typical value $\sim0.1$ and $\chi_{\rm K}^{\rm N}\equiv\partial n_{\rm K}/\partial \mu_{\rm K}$ is evaluated at nuclear matter in equilibrium (\rm N). Both $k_{\rm F}^{\Sigma}$ and $\chi_{\rm K}^{\rm N}$ are functions of $n_{\rm N}$, depending on the nuclear matter EoS. The timescale for  relevant weak interactions to happen in nuclear matter is much longer   than that in quark matter [Eq.~\eqn{eqn:tau_qm}], because the baryons are non-relativistic and their densities are lower. For typical transition densities in hybrid stars we studied, the ratio $\tau_{\rm N}/\tau_{\rm Q}$ is of order $10^{2}$.

To estimate the diffusion coefficient $D_{\rm N}\simeq\frac{1}{3}v_{\rm N}\lambda_{\rm N}$ we estimate $v_{\rm N}$ by the Fermi velocity of hyperons ($v_{\rm N}\simeq v_{\rm F}^{\Sigma}=k_{\rm F}^{\Sigma}/m_{\rm \Sigma}^{\ast}$), and the mean free path by $\lambda_{\rm N}\simeq v_{\rm F}^{\Sigma}/\nu_{\rm n\Sigma}$, where $\nu_{\rm n\Sigma}$ is the hadron-hyperon collision frequency similar to the hadron-hadron collision frequency $\nu_{\rm np}$ (see Eq.~(55) of Ref.~\cite{Shternin:2008es}). As a result,
\beq
D_{\rm N} \simeq \frac{m_{\rm n}^2\,{k_{\rm F}^{\Sigma}}^{2}}{32\,m_{\rm n}^{\ast}\,m_{\rm \Sigma}^{\ast 4}\,T^2\,S_{\rm n\Sigma}(k_{\rm F}^{n}, k_{\rm F}^{\Sigma})},
\label{eqn:D_nm}
\eeq
where $S_{\rm n\Sigma}$ is the effective hadron-hyperon scattering cross section which we for simplicity approximate by the proton-neutron cross section given in Eq.~(58) of Ref.~\cite{Shternin:2008es}. 
Given the nuclear matter EoS, $k_{\rm F}^{\Sigma}$ and $S_{\rm n\Sigma}$ can be expressed in terms of the baryon density $n_{\rm N}$. The strangeness diffusion coefficient in nuclear matter is typically much smaller than in quark matter [Eq.~\eqn{eqn:D_qm}], because hadrons and hyperons are non-relativistic while quarks are moving nearly at the speed of light, and also because the long-range interactions between the quark interactions give the different temperature dependence $D_{\rm N}/D_{\rm Q}\propto (T/\mu)^{1/3}$. At temperatures relevant to neutron stars, this ratio is of order $10^{-2}$.

The $K$-fraction in nuclear matter at the boundary with quark matter, $b_{\rm N^{\star}}$, is constrained by how far out of equilibrium the boundary is. For the boundary to move towards its ``ideal'' position the total chemical potential per baryon number in $\beta$-equilibrated quark matter ($\rm Q$ in Fig.~\ref{fig:nm-qm}) must be larger than that in the non-$\beta$-equilibrated nuclear matter on the other side of the boundary ($\rm N^{\star}$). Following the same logic as in Sec.~\ref{sec:nm_to_qm} we obtain a condition similar to Eq.~\eqn{eqn:dmu_a0}: 
\beq
\mu_{\rm B}^{\rm Q}-\mu_{\rm B}^{\rm N}>\frac{\left(-\mu_{\rm K}^{\rm N^{\star}}\right)^2\chi_{\rm K}^{\rm N}}{2n_{\rm N}}.
\label{eqn:dmu_b0}
\eeq
From Eq.~\eqn{eqn:QtoN-BC}, $b_{\rm N^{\star}}=-n_{\rm K}^{\rm N^{\star}}/n_{\rm N}\approx-\mu_{\rm K}^{\rm N^{\star}}\chi_{\rm K}^{\rm N}/n_{\rm N}$, so Eq.~\eqn{eqn:dmu_b0} gives an upper bound on $-b_{\rm N^{\star}}$ and hence on the front speed $v_{\rm {Q\to N}}$
\beq
(-b_{\rm N^{\star}})^{\rm max}=\sqrt{\frac{2\Delta\mu_{\rm B}\chi_{\rm K}^{\rm N}}{n_{\rm N}}},
\label{eqn:b0_max}
\eeq
\beq
\Delta\mu_{\rm B}=\mu_{\rm B}^{\rm Q}-\mu_{\rm B}^{\rm N}\simeq(\gamma-1)\delta p/n_{\rm Q}, 
\label{eqn:dmu_dp_nm}
\eeq
where
\beq
\delta p=g_{\rm b}\ep_{\rm crit}^{\rm N}(\delta x_{\rm b}(t)-\Delta x_{\rm ib}\sin(\omega t))\ .
\label{eqn:dp_dx_nm}
\eeq
Assuming that the boundary moves at its maximum speed, $(-b_{\rm N^{\star}})\approx (-b_{\rm N^{\star}})^{\rm max}$, the boundary velocity in the $\rm QM \to NM$ half cycle ($t_1<t<t_2$ in Fig.~\ref{fig:x_ideal_real}) is
\beq
\frac{\mathrm {d} \delta x_{\rm b}}{\mathrm {d}t}
\simeq \frac{1}{b_{\rm Q}}\sqrt{\frac{D_{\rm N}}{2\tau_{\rm N}}}\sqrt{\bigl[\de z /\ell_{\rm N} \bigr]^2+2\eta_{\rm N}\de z /\ell_{\rm N}}
\label{eqn:dx_t_nm} 
\eeq
where $\de z \equiv |\delta x_{\rm b}-\delta x_{\rm ib}|$ is how far the boundary
is from its equilibrium position at the current pressure, with the typical length
\beq
\ell_{\rm N}= \frac{(n_{\rm N}/\chi_{\rm K}^{\rm N})n_{\rm Q}}{2(\gamma-1)g_{\rm b}\ep_{\rm crit}^{\rm N}} \ .
\label{eqn:a_tilde_nm}
\eeq

Therefore with the periodic condition $\delta x_{\rm b}(t)=\delta x_{\rm b}(2\pi/\omega+t)$, Eqs.~\eqn{eqn:dx_t_qm} and \eqn{eqn:dx_t_nm} fully specify the movement of the phase boundary in response to the external pressure oscillation. Next we compute the energy dissipation in this process and see whether it is capable of saturating the r-mode.

\subsection{Dissipated power and saturation amplitude}
\label{sec:v_discussion}

From Sec.~\ref{sec:r-mode-diss} we know that during one cycle of an r-mode of amplitude $\al$ the energy dissipated  in a radially oriented cylinder
with an infinitesimal base area $\mathrm{d}S$ straddling the phase
boundary at $(\th,\phi)$ is
\beq
 \mathrm{d}W(\alpha, \theta, \phi)=\mathrm{d}S\left(\gamma-1\right)\Delta p_{\rm N} \int_0^\tau\!\!\cos(2\phi+\omega t)\frac{\mathrm{d}\delta R_{\rm b}}{\mathrm{d}t} \mathrm{d}t,
\label{eqn:dW_dp}
\eeq
where the position of the phase boundary $\delta R_{\rm b}(t)$ is the same as $\delta x_{\rm b} (t)$ in Secs.~\ref{sec:nm_to_qm} and \ref{sec:qm_to_nm}, which we assume to move at its maximal speed [see Eqs.~(\ref{eqn:dx_t_qm}) and (\ref{eqn:dx_t_nm})], and $\mathrm {d}S=\bar{R}_{\rm b}^2\sin\theta\mathrm {d}\theta\mathrm {d}\phi$. From Eq.~\eqn{eqn:dpL_rmode}
\beq
\Delta p_{\rm N}=g_{\rm b}\ep_{\rm crit}^{\rm N}|\delta R_{\rm ib}|=\ep_{\rm crit}^{\rm N}C_{\rm b}\alpha|\sin^2\!\theta\cos \theta| \, 
\label{eqn:dpN_rmode}
\eeq

Integrating Eq.~\eqn{eqn:dW_dp} over the full range of solid angle gives the total
dissipation of the r-mode in one cycle of oscillation and hence
the total power dissipated, $P_{\rm dis}$.
The r-mode amplitude stops growing (saturates) when this equals
the power injected via back-reaction from gravitational
radiation $P_{\rm gr}$.

\begin{figure*}
\parbox{0.5\hsize}{
\vspace{-2ex}
\includegraphics[width=\hsize]{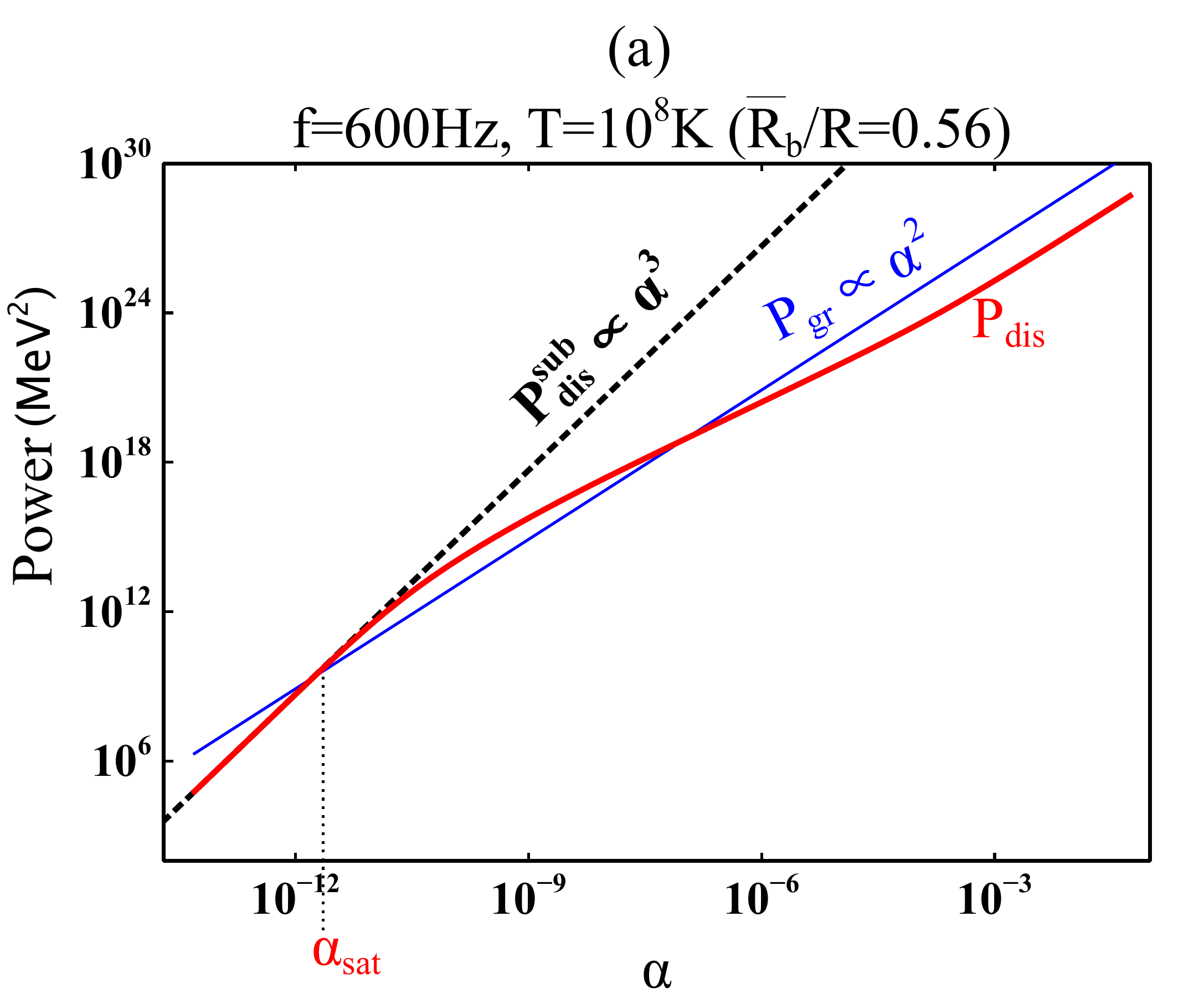}\\[2ex]
}\parbox{0.5\hsize}{
\vspace{-2ex}
\includegraphics[width=\hsize]{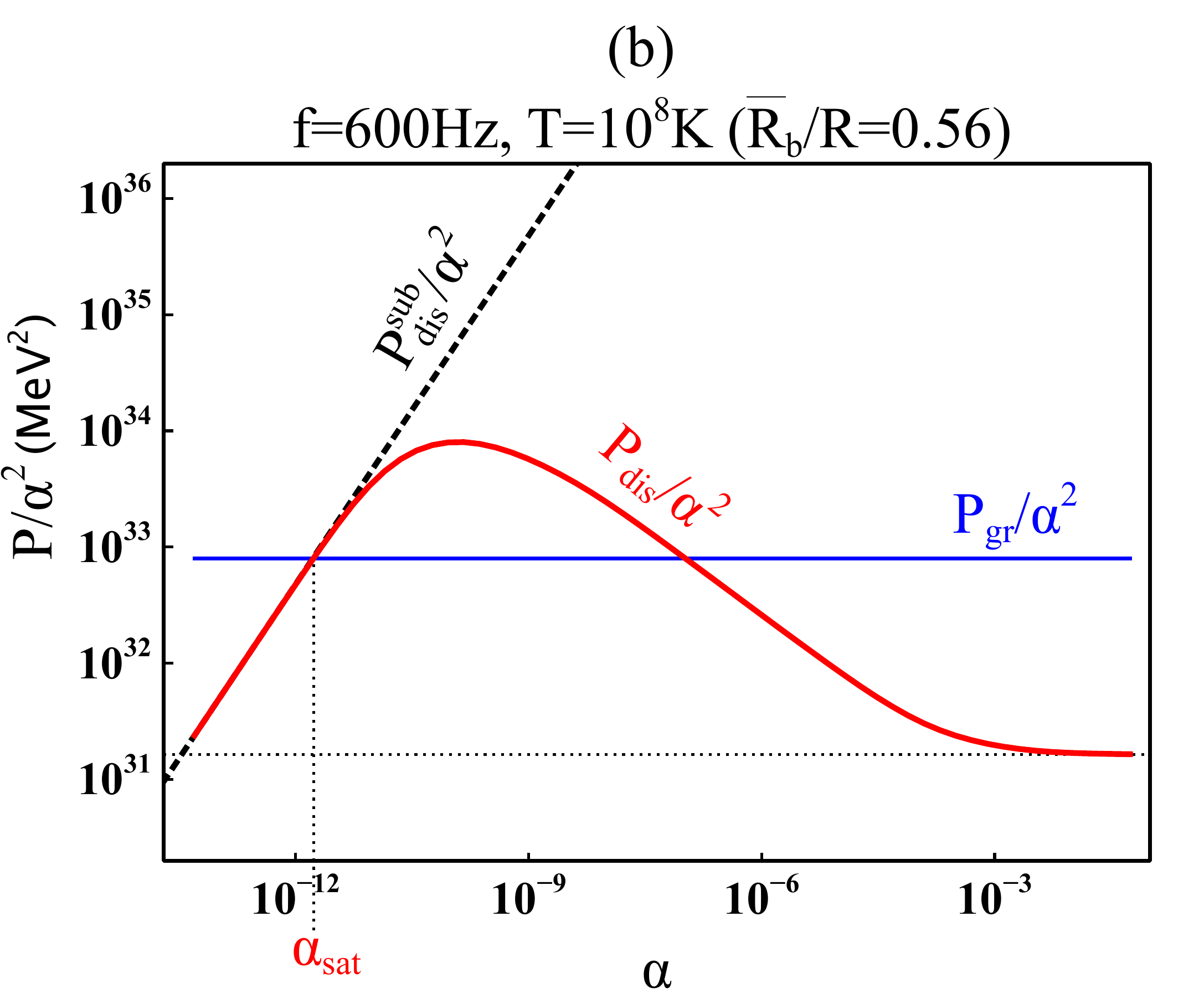}\\[2ex]
}
\caption{(Color online) (a) Dissipated power due to phase conversion $P_{\rm dis}$ (thick solid red [gray] curve) as a function of r-mode amplitude $\al$ for a specific example hybrid star (see text). (b) The same quantity where the vertical axis now shows the ratio $P_{\rm dis}/\alpha^2$. At first $P_{\rm dis}$ is proportional to $\alpha^3$ at very low amplitude (dashed line), then at some intermediate amplitude varies less quickly, with a maximum in $P_{\rm dis}/\al^2$, and finally changes to $\al^2$ at higher amplitude.
Also shown is gravitational radiation power $P_{\rm gr}$ (thin solid blue [gray] straight line), which is proportional to $\al^2$ at all amplitudes.
The r-mode amplitude will stop growing when dissipation balances
radiation, at the first point of intersection between the two curves.
This defines the saturation amplitude $\alpha_{\rm sat}$.}
\label{fig:P_alpha_complete_HS0}
\end{figure*} 

As an illustrative example, Fig.~\ref{fig:P_alpha_complete_HS0} 
shows the dissipated power as a function of r-mode amplitude
for a hybrid star rotating with frequency $f=600\,{\rm Hz}$, with 
quark core size $\bar{R}_{\rm b}/R=0.56$
and temperature $T=10^8\,{\rm K}$. For the quark matter EoS
we use the ``constant speed of sound'' (CSS) parametrization \cite{Alford:2013aca} 
with $n_{\rm trans}=4 n_0$, $\Delta \ep/\ep_{\rm trans}=0.2$, and 
$c_{\rm QM}^2=1$. The hadronic matter EoS is taken from 
Ref.~\cite{Hebeler:2010jx}.

In the subthermal regime, the dissipated power first rises with the r-mode amplitude $\al$ as $\al^3$ at very low amplitude, before entering a resonant region with a maximum in $P_{\rm dis}/\al^2$. At high amplitude in the suprathermal regime, the dissipated power is proportional to $\al^2$. The power in gravitational radiation from
the r-mode $P_{\rm gr}$ [Eq.~\eqn{eqn:dedt_rmode}] rises as $\al^2$, and is also shown in Fig.~\ref{fig:P_alpha_complete_HS0} for this particular hybrid star.
At low amplitude, the phase conversion dissipation is suppressed
relative to the gravitational radiation and therefore plays no role 
in damping the r-mode. If other damping mechanisms are too
weak to suppress the r-mode, its amplitude will grow. However,
as the amplitude grows, the phase conversion dissipation becomes stronger, and in this example there is a saturation amplitude $\al_{\rm sat}$ at which it equals the gravitational radiation, and the mode stops growing.

Varying parameters such as the size of the quark matter core,
rotation frequency, or temperature of the star will shift the curves in
Fig.~\ref{fig:P_alpha_complete_HS0}, and if the phase conversion dissipation is
too weak then there will be no intersection point ($P_{\rm gr}$ will be
greater than $P_{\rm dis}$ at all $\al$) and phase conversion
dissipation will not stop the growth of the mode.
However, we can see from Fig.~\ref{fig:P_alpha_complete_HS0} that
if saturation occurs, the resultant $\al_{\rm sat}$ is in
the low-amplitude regime, where an analytical approach is available,
and the saturation amplitude is extraordinarily low, of order $10^{-12}$.
This is typical of all model hybrid stars that we investigated.
In Appendix \ref{app:subthermal}
we derive the analytical expression for the dissipated power 
in the low-amplitude regime (dashed [black] line in Fig.~\ref{fig:P_alpha_complete_HS0}), obtaining
\beq
P_{\rm dis}^{\rm sub}(\alpha)\approx \frac{\alpha^3}{15}\left(\frac{105}{756\pi}\right)^{3/2}\frac{\gamma-1}{\Delta \tilde{p}_{\rm N}}\frac{(\ep_{\rm crit}^{\rm N})^2 \Omega^7 \bar{R}_{\rm b}^{11}}{g_{\rm b} R^3} \ .
\label{eqn:P_sub_approx}
\eeq
This expression allows us to assess how the strength of
phase conversion dissipation depends
on the various parameters involved.
It is particularly sensitive to the size of the quark core, 
and this will be important when
considering a whole family of hybrid stars with different
central pressures and hence different core sizes.

\begin{figure}[htb]
\includegraphics[width=\hsize]{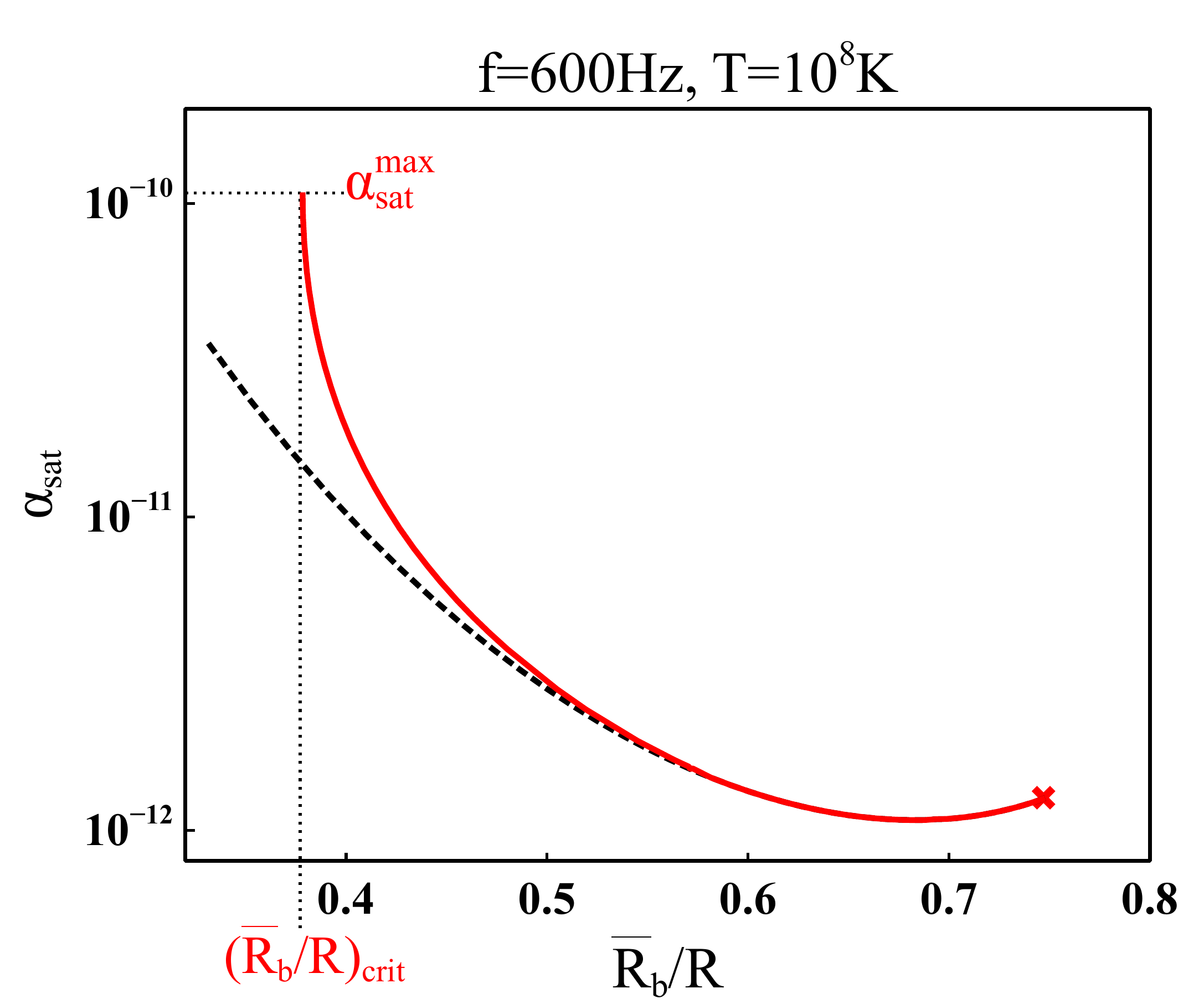}
\caption{(Color online) R-mode saturation amplitude (red [gray] solid curve) and its low-amplitude analytical approximation (black dashed curve) as a function of the 
radius of the quark matter core $\bar{R}_{\rm b}$ divided by the star radius $R$ in a family of hybrid stars.
For $\bar{R}_{\rm b}/R < (\bar{R}_{\rm b}/R)_{\rm crit}\approx0.38$, damping is too weak
to saturate the r-mode. At $\bar{R}_{\rm b}/R \gtrsim 0.75$ the
hybrid star is unstable against gravitational collapse. The mass fraction of the core is in the range $0.12\lesssim M_{\rm core}/M_{\rm star}\lesssim0.68$ for all the configurations shown on the red solid curve.}
\label{fig:alpha_sat_core_ratio_HS0_600Hz}
\end{figure}

The results of such an investigation are shown in
Fig.~\ref{fig:alpha_sat_core_ratio_HS0_600Hz}, where the solid (red online) curve
gives the numerically calculated saturation amplitude
($\alpha_{\rm sat}$ in Fig.~\ref{fig:P_alpha_complete_HS0}) as a function of
the size of the quark matter core in units of the star radius,
$\bar{R}_{\rm b}/R$. To construct this curve we used
the hadronic and quark matter EoSes of
Fig.~\ref{fig:P_alpha_complete_HS0} and varied the central
pressure, yielding a family of different star configurations. As
$\bar{R}_{\rm b}/R$ decreases, the dissipation power $P_{\rm dis}$ 
decreases rapidly relative to the gravitational radiation $P_{\rm gr}$.
The relative shift in the two corresponding curves in
Fig.~\ref{fig:P_alpha_complete_HS0} leads to
an upper limit on $\alpha_{\rm sat}$ when
$P_{\rm gr}$ is tangent to $P_{\rm dis}$.
This corresponds to the end of the solid curve 
in Fig.~\ref{fig:alpha_sat_core_ratio_HS0_600Hz} at
$\alpha_{\rm sat}^{\rm max}$ at the
critical value of the quark core size, $(\bar{R}_{\rm b}/R)_{\rm crit}$, below
which the phase-conversion mechanism cannot saturate the r-mode any more.

The black dashed curve in
Fig.~\ref{fig:alpha_sat_core_ratio_HS0_600Hz} is the low-amplitude analytical approximation to $\al_{\rm sat}$ [see also Appendix~\ref{app:subthermal}, Eq.~\eqn{eqn:alpha_sat_approx}],
\bea
&&\alpha_{\rm sat}^{\rm approx}=\left(\frac{2^{22}\pi^{9/2}}{3^3\cdot5^{5/2}} \right)\!G\, \frac{\tilde{D}_{\rm N}}{\tau _{\rm N}} \frac{\left(\chi_{\rm K}^{\rm N}\right)^3}{n_{\rm Q}n_{\rm N}^3 b_{\rm Q}^2} \frac{g_{\rm b}^3 M^2 \tilde{J}^2}{\Omega} \frac{R^9}{\bar{R}_{\rm b}^{11}}\nonumber \\[2ex]
&&\approx 4.2\times 10^{-11}\;\gamma \left(\frac{\tilde{D}_{\rm N}}{1.5\,{\rm MeV}}\right)\left(\frac{\tau_{\rm N}}{2\times 10^{-8}\,{\rm s}}\right)^{-1}\nonumber \\
&&\times\  \left(\frac{b_{\rm Q}}{1/3}\right)^{-2} \left(\frac{n_{\rm N}}{2\,n_0}\right)^{-4} \left(\frac{\chi_{\rm K}^{\rm N}}{(100\,{\rm MeV})^2}\right)^{3}\left(\frac{g_{\rm b}}{g_{\rm u}}\right)^{3}\nonumber \\
&&\times \left(\frac{\ep_{\rm crit}^{\rm Q}}{2\,\ep_{\rm crit}^{\rm N}}\right)^{3}\left(\frac{\ep_{\rm crit}^{\rm N}}{600\,{\rm MeV\,fm^{-3}}}\right)^{3}  \left(\frac{M}{1.4\,M_{\odot}}\right)^{2}\nonumber \\
&&\times  \left(\frac{\tilde{J}}{0.02}\right)^2 \left(\frac{f}{1 {\rm kHz}}\right)^{-1} \left(\frac{R}{10\,{\rm km}}\right) \left(\frac{\bar{R}_{\rm b}/R}{0.4}\right)^{-8},
\label{eqn:alpha_sat_analytic}
\eea
where $D_{\rm N}\equiv \tilde{D}_{\rm N}\times T^{-2}$ [see Eq.~\eqn{eqn:D_nm}], and $g_{\rm u}$ is the Newtonian gravitational acceleration at the phase boundary when the quark core has uniform density $\ep=\ep_{\rm crit}^{\rm Q}$ ($g_{\rm u}\equiv \frac{4}{3}\pi G \ep_{\rm crit}^{\rm Q} \bar{R}_{\rm b}$).
This approximation is very accurate when the phase conversion damping
is strong, but it does not capture the sudden weakening of that dissipation
when, for example, the core radius becomes small. It is therefore useful to
have some idea of its range of validity.

\subsection{Range of validity of low-amplitude approximation}

The range of validity of 
Eqs.~(\ref{eqn:P_sub_approx}) and (\ref{eqn:alpha_sat_analytic})
is found by calculating the next-to-leading (NLO) contribution, and requiring that it be less than a fraction $\epsilon$ of the total dissipated power.
We find (see Appendix~\ref{app:validity}) that the approximation is valid when
\bea
&& \epsilon \geqslant  \frac{2^{39}\pi^5}{3^{12}5^4} \frac{G^2}{(\gamma-1)^2}\frac{g_{\rm b}^2 M^4 \tilde{J}^4\Omega^6}{\left(\ep_{\rm crit}^{\rm N}\right)^2}\left(\frac{R}{\bar{R}_{\rm b}}\right)^{16}\nonumber \\
&&\simeq 2.96 \left(\frac{\gamma-1}{0.5}\right)^{-2} \left(\frac{\ep_{\rm crit}^{\rm Q}}{2\,\ep_{\rm crit}^{\rm N}}\right)^{2} \left(\frac{g_{\rm b}}{g_{\rm u}}\right)^{2} \left(\frac{M}{1.4\,M_{\odot}}\right)^{4} \nonumber \\
&&\times \left(\frac{\tilde{J}}{0.02}\right)^{4} \left(\frac{f}{1\,\rm{kHz}}\right)^{6}\left(\frac{R}{10\,{\rm km}}\right)^{2}\left(\frac{\bar{R}_{\rm b}/R}{0.4}\right)^{-14} \!\!  .
\label{eqn:ep_astro}
\eea
We see that the validity of the low-amplitude approximation is
mainly determined by the size of the quark matter core and the
rotation frequency of the star.

\section{Conclusion}
\label{sec:conclusion}

We have described how phase conversion
in a multicomponent compact star provides a mechanism for
damping density oscillations, via the 
phase lag in the response of the interface between components
of different baryon densities to the applied pressure oscillation.
The phase lag arises from the finite rate of interconversion between
the phases, which limits the speed with which the interface can move.
We studied the case where the two phases are separated by a 
sharp boundary (first-order phase transition)
and analyzed the movement of the interface in the approximation
of a steady state, neglecting additional acceleration effects and
complicated hydrodynamic effects like
turbulence. In particular, we studied the astrophysically
interesting case of the damping of r-mode oscillations
\cite{Andersson:1997xt,Andersson:2000mf} in a two-component star.
We found that  phase conversion dissipation does not affect the
r-mode instability region, because it vanishes as $\al^3$ at low
r-mode amplitude $\al$. 
However, depending on the values of relevant parameters,
phase conversion dissipation can either saturate the r-mode at extremely low amplitudes,
$\al_{\rm sat} \lesssim 10^{-10}$ 
in the explicit example of hadron-quark transformation
at the sharp quark-hadron interface in a hybrid star, or be insufficient to saturate the r-mode at all.
The reason for this behavior stems, analogously to the bulk visocity \cite{Alford:2010gw}, from the resonant character of the dissipation, which is relatively strong when the time scale of the dissipation matches the time scale of the external oscillation (see Fig.~\ref{fig:P_alpha_complete_HS0}). Whether saturation is possible depends therefore on the microscopic and astrophysical parameters, like in particular on the mass of the quark core, which should not be too small.

Our main result is Eq.~\eqn{eqn:dW_dp}, which must be evaluated using numerical solutions of Eqs.~\eqn{eqn:dx_t_qm} and \eqn{eqn:dx_t_nm}. We also give the low-amplitude analytic expressions for the power dissipated
[Eq.~\eqn{eqn:P_sub_approx}] and the saturation amplitude 
[Eq.~\eqn{eqn:alpha_sat_analytic}] which are valid
when the dissipation is sufficiently strong, obeying Eq.~\eqn{eqn:ep_astro}
with $\epsilon \ll 1$.

Our results have significant implications for astrophysical
signatures of exotic high-density phases of matter, such as quark matter.
The observed data for  millisecond pulsars is not consistent with
the minimal model of pulsars as stars made of nuclear matter
with damping of r-modes
via bulk and shear viscosity \cite{Alford:2013pma}. Resolving this discrepancy
requires either a new mechanism for stabilizing r-modes, or
a new mechanism for saturating unstable r-modes at 
$\alpha_{\rm sat} \lesssim 10^{-8} \!\!-\!\!10^{-7}$ \cite{Mahmoodifar:2013quw,Alford:2013pma,Haskell:2012vg}. Previously proposed mechanisms have problems to achieve this.
Suprathermal bulk viscosity and hydrodynamic oscillations 
both give $\al_{\rm sat}\sim 1$ \cite{Alford:2011pi,Lindblom:2000az}.
The nonlinear coupling of the r-mode to viscously
damped daughter modes could give $\alpha_{\rm sat} \sim 10^{-6}$ to $10^{-3}$ 
\cite{Brink:2004bg,Bondarescu:2013xwa}. 
The recently proposed vortex-fluxtube cutting mechanism
\cite{Haskell:2013hja} might give sufficiently small 
saturation amplitudes but is present only at sufficiently low temperatures $T \ll T_{c}\lesssim 10^9$ K, which could be exceeded by the r-mode (and/or accretion) heating \cite{Alford:2013pma}.
One of the main results of this paper is that
phase conversion dissipation can provide saturation at the required
amplitude to explain millisecond pulsar data.

Second, due to the extremely low r-mode saturation amplitude of our proposed mechanism, hybrid stars would behave very differently from neutron or strange stars. As discussed in Ref.~\cite{Alford:2013pma}, if the known millisecond sources were hybrid stars then, for the low saturation amplitudes that we have found, they would have cooled out of the r-mode instability region quickly (in millions of years) so that they would have very low temperatures by now. In contrast, in neutron stars r-modes would be present and would provide such strong heating that the temperature of observed millisecond pulsars would be $T_\infty \sim O(10^5\!\!-\!\!10^6)$ K \cite{Alford:2013pma}. This prediction assumes a (so far unknown) saturation mechanism that would saturate the mode at a value $\alpha_{\rm sat} \lesssim 10^{-8}$ required by the pulsar data. This temperature is significantly higher than what standard cooling estimates suggest for such old sources. The same holds for strange quark stars where the enhanced viscous damping can explain the pulsar data, but even in this case the star would spin down along the boundary of the corresponding stability window which would keep it at similarly high temperatures. Measurements of or bounds on temperatures of isolated millisecond pulsars provide therefore a promising way to discriminate hybrid stars.

Our analysis considered only strangeness-changing non-leptonic processes when
we discussed the hadron-quark transformation as an example of
phase conversion dissipation. 
However, there are also leptonic processes that equilibrate the
non-strange neutron-proton or up-down ratio. For ordinary 
bulk viscosity in hadronic or quark matter these processes are
only relevant at temperatures far above the temperature of a neutron star
because their rate is
parametrically smaller than the strangeness changing rate discussed here 
by a factor of $(T/\mu)^2$ \cite{Haensel:1992zz}. 
However, leptonic processes might play an important role in phase conversion
dissipation because hadronic matter has more electrons and up quarks than
quark matter, so, just as for strangeness, there will be a conversion region
behind the moving boundary where conversion and diffusion of up-ness
is occurring. Taking this into account could  change the estimates given here and should be studied in more detail in the future.

As well as the quark-hadron interface in a hybrid star, any first-order phase
transition that leads to a sharp interface between two phases 
with different baryon densities could, via the
mechanism discussed here, cause dissipation of global pressure oscillation
modes. 
One possibility would be different phases of quark matter, perhaps with
different Cooper pairing patterns, 
such as the color-flavor locked (CFL) phase,
the two-flavor color superconductor (2SC), or various forms of inhomogeneous and
asymmetric pairing \cite{Alford:2007xm}, which are all generally connected by
first-order phase transitions.  Because cross-flavor pairing induces shifts in
the Fermi surfaces of the participating species, different color
superconducting phases will often have different flavor fractions, so movement
of the interface between them requires weak interactions, as in the case of
the quark-hadron interface. The dissipation mechanism discussed here may
therefore be
expected to operate, albeit mildly suppressed by the smallness
of the baryon number density differences between these phases.

Our discussion was limited to the case of a sharp interface, which is the
expected configuration if the surface tension is large enough. If the surface
tension is small, there will instead be a mixed phase region where domains
of charged hadronic and quark matter coexist
\cite{Ravenhall:1983uh,Glendenning:1992vb}. We expect that
the phase conversion dissipation mechanism will operate in this case too,
as the domains expand and shrink in response to pressure oscillations.
However, to estimate this contribution is far more complicated since it
requires us to consider the dynamic formation, growth, and merging of 
these structures, taking into account the costs and
gains due to surface tension and electric field energy. Such an analysis is far
beyond the scope of this work, but we expect that the dissipation due to such
transformations will be roughly comparable to the estimates given here.
A similar mechanism
should also be relevant for the ``nuclear pasta'' mixed
phases in the inner crust of an ordinary neutron star. 
In this case in addition to the slow $\beta$-equilibration processes
there may also be slow strong interaction equilibration processes,
whose rate is suppressed by tunneling factors for the transition
between geometric domains of different size.
This could further enhance the dissipation.

The phase conversion mechanism for damping relies on the transition between
two phases being first order. If there is a crossover, then dissipation
due to particle conversion is described by the standard bulk viscosity.
Examples are the appearance of hyperons in the dense interior or the crossover from $npe$ to $npe\mu$ hadronic matter,
where the conserved particle
density is lepton number instead of baryon number \cite{Alford:2010jf}. 
The conversion is then not restricted to a thin transition region and
partial conversion, giving rise to bulk viscosity dissipation taking place all over the relevant part of the star. The additional effect, that the size of the region where muons are present changes as well, is negligible, since the muon fraction continuously goes to zero. This is also reflected by the vanishing of the prefactor in the parenthesis of our general expressions Eq.~(\ref{eq:W1_W2}).

In this work, we obtained a reasonable first estimate of the size of the damping 
by treating the movement of the phase boundary in the steady-state approximation \cite{Olinto:1986je}, 
assuming that it can accelerate arbitrarily fast and that
it can move as fast as allowed by general thermodynamic constraints.
In reality the acceleration of the phase boundary near the turning points
of its motion might be further slowed down by the fact that the steady-state conversion region has to form, and if it therefore cannot accelerate fast enough there will
be additional dissipation during this part of the cycle, even if the phase boundary is eventually fast enough to stay in chemical equilibrium near the equilibrium position. 
Our analysis showed that 
even being  out of chemical equilibrium for only 
a small fraction of a cycle causes the system to dissipate a huge amount of energy, so it is possible that
including these additional acceleration effects may yield an even lower
r-mode saturation  amplitude and saturate r-modes even in stars with small quark cores.
Including the realistic acceleration of the phase boundary will require solving the full time-dependent evolution of the phase conversion front. Similarly, it is likely that turbulence plays a major role in the phase conversion, as found in several analyses \cite{Pagliara:2013tza, Herzog:2011sn, Niebergal:2010ds} of the one-time burning of a (meta-stable) neutron star. The inclusion of these complications is an interesting future project.

\appendix
\section{Angular integral and saturation amplitude in the subthermal regime}
\label{app:subthermal}
To determine the velocity of the boundary in the $\rm {NM \to QM}$ half cycle, we define a dimensionless parameter $y(\varphi)\equiv \delta x_{\rm b}/\Delta x_{\rm ib}$ with $\varphi=\omega t$, and then Eq.~\eqn{eqn:dx_t_qm} becomes 
\beq
\left(\frac{\mathrm {d} y}{\mathrm {d}\varphi}\right)_{\rm N\to Q}
\simeq A_{\rm Q}\sqrt{\rho_{\rm Q}(\sin\varphi-y)^2+(\sin\varphi-y)},
\label{eqn:dy_phi_qm} 
\eeq
where $A_{\rm Q}$ represents an overall amplitude of the speed 
\beq
A_{\rm Q}= \frac{3}{a_{\rm N}}\sqrt{\frac{D_{\rm Q}}{2\tau_{\rm Q}}\frac{(\gamma-1)\eta_{\rm Q}}{(n_{\rm Q}/\chi_{\rm K}^{\rm Q})n_{\rm Q}}}\,\frac{g_{\rm b}\ep_{\rm crit}^{\rm N}}{\Omega\sqrt{\Delta p_{\rm N}}},
\label{eqn:A_qm} 
\eeq
while $\rho_{\rm Q}$ is the ratio of suprathermal to subthermal contribution 
\beq
\rho_{\rm Q}=\frac{(\gamma-1)\Delta p_{\rm N}}{\eta_{\rm Q}(n_{\rm Q}/\chi_{\rm K}^{\rm Q})n_{\rm Q}}.
\label{eqn:rho_qm} 
\eeq
Similarly in the $\rm {QM \to NM}$ half cycle Eq.~\eqn{eqn:dx_t_nm}  becomes 
\beq
\left(\frac{\mathrm {d} y}{\mathrm {d}\varphi}\right)_{\rm Q\to N}
\simeq A_{\rm N}\sqrt{\rho_{\rm N}(\sin\varphi-y)^2+(\sin\varphi-y)}
\label{eqn:dy_phi_nm} 
\eeq
and the two coefficients are
\bea
A_{\rm N}&=& \frac{3}{b_{\rm Q}}\sqrt{\frac{D_{\rm N}}{2\tau_{\rm N}}\frac{(\gamma-1)\eta_{\rm N}}{(n_{\rm N}/\chi_{\rm K}^{\rm N})n_{\rm Q}}}\,\frac{g_{\rm b}\ep_{\rm crit}^{\rm N}}{\Omega\sqrt{\Delta p_{\rm N}}}, \ \\
\label{eqn:A_nm} 
\rho_{\rm N}&=&\frac{(\gamma-1)\Delta p_{\rm N}}{\eta_{\rm N}(n_{\rm N}/\chi_{\rm K}^{\rm N})n_{\rm Q}}.
\label{eqn:rho_nm} 
\eea
Combining Eqs.~\eqn{eqn:dy_phi_qm} and \eqn{eqn:dy_phi_nm}, and setting $s(\varphi)\equiv\sin\varphi-y(\varphi)$ we have
\beq
\cos\varphi-\frac{\mathrm {d} s}{\mathrm {d}\varphi} =  \left\{\!
\begin{array}{lll}
A_{\rm N\to Q}\sqrt{\rho_{\rm Q} s(\varphi)^2+s(\varphi)},& s(\varphi)\geqslant 0, \\[1 ex]
A_{\rm Q\to N}\sqrt{\rho_{\rm N} s(\varphi)^2-s(\varphi)}, & {\rm otherwise.}
\end{array}
\right.
\label{eqn:dy_dphi_both}
\eeq
With the periodic condition ($s(\varphi)=s(2\pi+\varphi)$), one can solve for the full profile of the interface position and continue to calculate the dissipated energy.

The total
dissipation of the r-mode in one cycle of the oscillation is
\beq
W(\alpha)=\int\limits_0^\pi\!\int\limits_0^{2\pi}\mathrm {d}S (\gamma-1) \frac{(\Delta p_{\rm N})^2}{g_{\rm b} \ep_{\rm crit}^{\rm N}} V(\Delta p_{\rm N})
\label{eqn:W_full}
\eeq
where the integral $V(\Delta p_{\rm N})$ depends on the velocity of the
boundary in both directions
\beq
V(\Delta p_{\rm N})\equiv\int_0^{2\pi}\sin\varphi \;\frac{\mathrm{d}(\sin\varphi-s)}{\mathrm{d}\varphi} \;\mathrm{d}\varphi
\label{eqn:V_dp}
\eeq
and $s(\varphi)$ is the solution to Eq.~\eqn{eqn:dy_dphi_both}. In general this integral can be computed numerically as long as coefficients $A$'s and $\rho$'s in the differential equation Eq.~\eqn{eqn:dy_dphi_both} are known.

At small oscillation when $\Delta p_{\rm N}$ is tiny, however, the subthermal regime dominates ($\rho_{\rm Q}, \rho_{\rm N}\ll 1$) and Eq.~\eqn{eqn:dy_dphi_both} can be simplified as
\beq
\cos\varphi-\left(\frac{\mathrm {d} s}{\mathrm {d}\varphi}\right)_{\rm sub} =  \left\{\!
\begin{array}{lll}
A_{\rm Q}\sqrt{s(\varphi)},& s(\varphi)\geqslant 0, \\[1 ex]
A_{\rm N}\sqrt{-s(\varphi)}, & {\rm otherwise.}
\end{array}
\right.
\label{eqn:dy_dphi_sub}
\eeq
Since both $A_{\rm Q}$ and $A_{\rm N}$ are much greater than 1, to leading order the analytical solution to Eq.~\eqn{eqn:dy_dphi_sub} is
\beq
s(\varphi)=\Theta(\cos\varphi)\frac{\cos^2\!\varphi}{A_{\rm Q}^2}-\Theta(-\cos\varphi)\frac{\cos^2\!\varphi}{A_{\rm N}^2},
\label{eqn:s_phi_approx}
\eeq
where $\Theta$ is the Heaviside step function. 
The integral Eq.~\eqn{eqn:V_dp} becomes
\begin{align}
V_{\rm sub}(\Delta p_{\rm N})&= \frac{4}{3}\left( \frac{1}{A_{\rm Q}^2}+\frac{1}{A_{\rm N}^2} \right) \nonumber \\
&=\frac{4}{3}\Delta p_{\rm N}\left( \frac{1}{\Delta \tilde{p}_{\rm Q}}+\frac{1}{\Delta \tilde{p}_{\rm N}} \right),
\label{eqn:V_dp_approx_both}
\end{align}
where 
\bea
\Delta \tilde{p}_{\rm Q}&\equiv&\frac{9}{a_{\rm N}^2}(\gamma-1)\frac{D_{\rm Q}}{\tau_{\rm Q}}\frac{\eta_{\rm Q}\left(g_{\rm b}\ep_{\rm crit}^{\rm N}\right)^2}{n_{\rm Q}\left(n_{\rm Q}/\chi_{\rm K}^{\rm Q}\right)}\frac{1}{\Omega^2}, \ \\
\Delta \tilde{p}_{\rm N}&\equiv&\frac{9}{b_{\rm Q}^2}(\gamma-1)\frac{D_{\rm N}}{\tau_{\rm N}}\frac{\eta_{\rm N}\left(g_{\rm b}\ep_{\rm crit}^{\rm N}\right)^2}{n_{\rm Q}\left(n_{\rm N}/\chi_{\rm K}^{\rm N}\right)}\frac{1}{\Omega^2}. 
\label{eqn:dp_qm_nm_tilde}
\eea

At sufficiently low oscillation amplitude , the integral $V_{\rm sub}(\Delta p_{\rm N})$ is dominated by the term with the smaller value of $A$. For the class of models we have analyzed, in general $A_{\rm Q}\gg A_{\rm N}\gg 1$ ($\Delta \tilde{p}_{\rm Q}\gg\Delta \tilde{p}_{\rm N}\gg \Delta p_{\rm N}$), because in nuclear matter diffusion is less efficient ($D_{\rm N}/D_{\rm Q}\approx O(10^{-2})$) and weak interactions take more time to proceed ($\tau_{\rm N}/\tau_{\rm Q}\approx O(10^2)$). Therefore the $\rm {QM \to NM}$ transition half cycle dominates the dissipation and Eq.~\eqn{eqn:V_dp_approx_both} becomes 
\beq
V_{\rm sub}(\Delta p_{\rm N})\xrightarrow{\Delta \tilde{p}_{\rm Q}\gg \Delta \tilde{p}_{\rm N}\gg \Delta p_{\rm N}} \frac{4}{3}\Delta p_{\rm N}/\Delta \tilde{p}_{\rm N}.
\label{eqn:V_dp_sub_nm}
\eeq
Performing the angular integral in Eq.~\eqn{eqn:W_full} gives the expression for dissipated power $P_{\rm dis}\equiv W\Omega/(2\pi)$ at low amplitude
\beq
P_{\rm dis}^{\rm sub}(\alpha)\approx \frac{\alpha^3}{15}\left(\frac{105}{756\pi}\right)^{3/2}\frac{\gamma-1}{\Delta \tilde{p}_{\rm N}}\frac{(\ep_{\rm crit}^{\rm N})^2 \Omega^7 \bar{R}_{\rm b}^{11}}{g_{\rm b} R^3}.
\label{eqn:P_sub_nm}
\eeq
The power emitted by the mode as gravitational radiation is \cite{Lindblom:1998wf}
\beq
P_{\rm gr} \equiv
\left( \frac{\mathrm{d}\tilde{E}}{\mathrm{d}t}\right)_{\rm GR}=\frac{2^{17}\pi}{3^8 5^2}\alpha^2GM^2R^6\tilde{J}^2 \Omega^8.
\label{eqn:dedt_rmode} 
\eeq
The radial integral constant is given by
\beq
\tilde{J}\equiv\frac{1}{MR^4}\int_0^R \ep(r)r^6 \mathrm{d}r
\label{eqn:J_rmode}
\eeq
and its typical value for hybrid stars is $\simeq2\times 10^{-2}$.

The saturation amplitude $\alpha_{\rm sat}$ is determined by the equation
\beq
P_{\rm dis}=\left( \frac{\mathrm{d}\tilde{E}}{\mathrm{d}t}\right)_{\rm GR}.
\label{eqn:alpha_sat}
\eeq
By solving Eq.~\eqn{eqn:alpha_sat} with Eqs.~\eqn{eqn:dp_qm_nm_tilde}--\eqn{eqn:J_rmode}, we obtain the low-amplitude approximation for $\alpha_{\rm sat}$
\beq
\alpha_{\rm sat}^{\rm approx}=\left(\frac{2^{22}\pi^{9/2}}{3^3\times5^{5/2}} \right)\!G\, \frac{\tilde{D}_{\rm N}}{\tau _{\rm N}} \frac{\left(\chi_{\rm K}^{\rm N}\right)^3}{n_{\rm Q}n_{\rm N}^3 b_{\rm Q}^2} \frac{g_{\rm b}^3 M^2 \tilde{J}^2}{\Omega} \frac{R^9}{\bar{R}_{\rm b}^{11}},
\label{eqn:alpha_sat_approx}
\eeq
where $D_{\rm N}\equiv \tilde{D}_{\rm N}\times T^{-2}$ [see Eq.~\eqn{eqn:D_nm}].

\section{Range of validity of the analytical approximation for the saturation amplitude}
\label{app:validity}
To give an estimate for the validity of the analytic expression for the saturation amplitude, we have to compute Eq.~\eqn{eqn:s_phi_approx} to NNLO,
\begin{align}
s(\varphi)&=\cos^2\!\varphi\left[\frac{\Theta(\cos\varphi)}{A_{\rm Q}^2}-\frac{\Theta(-\cos\varphi)}{A_{\rm N}^2}\right]\nonumber \\
&+(4\cos^2\!\varphi\sin^2\!\varphi)\left( \frac{1}{A_{\rm Q}^4}+\frac{1}{A_{\rm N}^4} \right) \nonumber \\
&+(-8\cos^4\!\varphi+20\cos^2\!\varphi\sin^2\!\varphi) \nonumber \\
&\times\left[\frac{\Theta(\cos\varphi)}{A_{\rm Q}^6}-\frac{\Theta(-\cos\varphi)}{A_{\rm N}^6}\right]+\cdots
\label{eqn:s_phi_A}
\end{align}
in order to obtain the NLO correction to Eq.~\eqn{eqn:V_dp_approx_both}
\begin{align}
V_{\rm sub}(\Delta p_{\rm N})= &\frac{4}{3}\left( \frac{1}{A_{\rm Q}^2}+\frac{1}{A_{\rm N}^2} \right) \nonumber \\
&-\frac{16}{5}\left( \frac{1}{A_{\rm Q}^6}+\frac{1}{A_{\rm N}^6} \right)+\cdots
\label{eqn:V_sub_A}
\end{align}
This yields the correction term to the dissipated power in the low-amplitude regime
\bea
P_{\rm dis}^{\rm sub}(\alpha)&=&P_0(\alpha)+P_1(\alpha)+\cdots
\eea
where $P_0$ is the previous result Eq.~\eqn{eqn:P_sub_nm} and the NLO correction reads
\bea
P_1(\alpha)&=&-\frac{2\alpha^5}{105}\left(\frac{105}{756\pi}\right)^{5/2}\nonumber \\
&&\times\frac{\gamma-1}{\left(\Delta \tilde{p}_{\rm N}\right)^3}\frac{(\ep_{\rm crit}^{\rm N})^4 \Omega^{11} \bar{R}_{\rm b}^{17}}{g_{\rm b}^3 R^5}.
\label{eqn: P_sub_alpha}
\eea
The requirement that the analytical approximation for the saturation amplitude [Eq.~\eqn{eqn:alpha_sat_approx}] deviates from the exact result by less than a fraction $\epsilon$; i.e., $P_1(\alpha_{\rm sat}^{\rm LO})\leqslant \epsilon P_0(\alpha_{\rm sat}^{\rm LO})$ yields the bound on the underlying parameters
\bea
\frac{2^{39}\pi^5}{3^{12}5^4} \frac{G^2}{(\gamma-1)^2}\frac{g_{\rm b}^2 M^4 \tilde{J}^4\Omega^6}{\left(\ep_{\rm crit}^{\rm N}\right)^2}\left(\frac{R}{\bar{R}_{\rm b}}\right)^{16} &\leqslant& \epsilon.
\label{eqn:P1_P0}
\eea
We can see for the set of parameter values on the left-hand side below $\epsilon \ll 1$, Eq.~\eqn{eqn:alpha_sat_approx} is a good approximation.
With the EoS we applied in Fig.~\ref{fig:alpha_sat_core_ratio_HS0_600Hz}, for $\bar{R}_{\rm b}/R=0.56$, the left-hand side is $\approx0.04$; for $\bar{R}_{\rm b}/R=(\bar{R}_{\rm b}/R)_{\rm crit}=0.38$ at $\alpha=\alpha_{\rm sat}^{\rm max}$, it is $\approx 4$.

\section*{ACKNOWLEDGMENTS}
We thank Mikhail Gusakov for pointing out the chemical potential constraint on front speeds and Charles Horowitz and Brynmor Haskell for useful discussions.
This research was supported in part by the 
Office of Nuclear Physics 
of the U.S.~Department of Energy under Contract
 No. DE-FG02-05ER41375 
 and by the DOE Topical Collaboration 
 ``Neutrinos and Nucleosynthesis in Hot and Dense Matter'', 
 Contract No. DE-SC0004955.

\appendix

\renewcommand{\href}[2]{#2}

\newcommand{\apjl}{Astrophys. J. Lett.\ }
\newcommand{\mnras}{Mon. Not. R. Astron. Soc.\ }
\newcommand{\aap}{Astron. Astrophys.\ }

\bibliographystyle{JHEP_MGA}
\bibliography{cs.bib} 

\end{document}